\documentclass[pdflatex,sn-mathphys-num]{sn-jnl}

\usepackage{graphicx}%
\usepackage{multirow}%
\usepackage{amsmath,amssymb,amsfonts}%
\usepackage{amsthm}%
\usepackage{mathrsfs}%
\usepackage[title]{appendix}%
\usepackage{xcolor}%
\usepackage{textcomp}%
\usepackage{manyfoot}%
\usepackage{booktabs}%
\usepackage{algorithm}%
\usepackage{algorithmicx}%
\usepackage{algpseudocode}%
\usepackage{listings}
\theoremstyle{thmstyleone}%
\theoremstyle{thmstyletwo}%

\theoremstyle{thmstylethree}%

\begin{document}

\title[Article Title]{Floquet analysis of coherence in periodically driven diamond  nitrogen-vacancy ensemble systems}


\author[1]{\fnm{Manh-Cuong} \sur{Nguyen}}

\author[1]{\fnm{Uijin} \sur{Ko}}

\author[2]{\fnm{Seong-Joo} \sur{Lee}}

\author[3]{\fnm{Hyeonsu} \sur{Kim}}

\author*[3,4]{\fnm{Hosung} \sur{Seo}}\email{seo.hosung@skku.edu}

\author*[1]{\fnm{Sangwon} \sur{Oh}}\email{sangwonoh@ajou.ac.kr}

\affil[1]{\orgdiv{Department of Physics and Energy Systems Research}, \orgname{Ajou University}, \orgaddress{ \city{Suwon}, \postcode{16499}, \country{Republic of Korea}}}

\affil[2]{\orgdiv{Quantum Magnetic Sensing Group}, \orgname{Korea Research Institute of Standards and Science}, \orgaddress{\city{Daejeon}, \postcode{34113}, \country{Republic of Korea}}}

\affil[3]{\orgdiv{SKKU Advanced Institute of Nanotechnology}, \orgname{Sungkyunkwan University}, \orgaddress{ \city{Suwon}, \postcode{16419}, \country{Republic of Korea}}}

\affil[4]{\orgdiv{Department of Quantum Information Engineering}, \orgname{Sungkyunkwan University}, \orgaddress{ \city{Suwon}, \postcode{16419}, \country{Republic of Korea}}}


\abstract{
	
	High-density nitrogen-vacancy (NV) ensembles are promising platforms for solid-state quantum sensing, but their performance is limited by dipolar interactions and inhomogeneous dephasing. Periodic decoupling sequences such as Waugh-Huber-Haeberlen (WAHUHA) can extend the observed stroboscopic decay time, but a longer fitted lifetime does not necessarily imply improved magnetic-field sensitivity. Here, we experimentally demonstrate this disconnect experimentally. WAHUHA increases the fitted stroboscopic decay time from $T_{2}^*$ of 0.9 $\mu$s to $T_{2,eff}^*$ of 31 $\mu$s, while producing little improvement in dc magnetic-field sensitivity. Using detuning-resolved stroboscopic spectroscopy and finite-pulse Floquet analysis, we show that the long-lived signal arises from phase wrapping and quasienergy branch folding of the one-cycle unitary. These effects reshape the stroboscopic spectrum, and the experimentally relevant folded branches exhibit a reduced local detuning-to-phase transduction slope, $d\Phi/d\Delta$, which governs the dc magnetic-field response. Our results demonstrate that, under periodic driving, an extended effective dephasing time does not necessarily translate into enhanced dc sensitivity and establish finite pulse Floquet analysis as a practical framework for evaluating coherence in spin ensembles.
	
}

\keywords{Floquet Analysis, Floquet Engineering, Coherence, Nitrogen-vacancy Ensemble, Quantum Sensing, Spin Dynamics, Multi-pulse Sequence}



\maketitle

\section{Introduction}\label{sec1}

Nitrogen-vacancy (NV) centers in diamond have emerged as a versatile solid-state platform for quantum sensing, owing to their optically addressable spin states, long coherence times, and room temperature operation \cite{ Degen2017,Janitz2022,Doherty2013,Jelezko2004a,Jelezko2004b}. In the single-defect regime, these properties have enabled nanoscale magnetic imaging \cite{Maletinsky2012,Chang2017,Hopper2018,Huxter2023,Keppler2022,Marchiori2022,Mosavian2024}, single-molecule nuclear magnetic resonance (NMR) and electron paramagnetic resonance (EPR) spectroscopy \cite{Staudacher2013,Mamin2013}, quantum information processing and communication \cite{Hensen2015,Wu2023}, and nanoscale thermometry in biological environments \cite{Kucsko2013}. Extending from single defects to ensembles allows a macroscopic number of spins to contribute to the optical signal, thereby improving the signal-to-noise ratio and enabling wide-field magnetic imaging \cite{Barry2020,LeSage2013,Cheng2025,Cambria2025}, bulk ultra-sensitive magnetometry \cite{Hughes2025,Wang2022,Marshall2022}, micro-scale NMR \cite{Glenn2018}, room-temperature masers \cite{Breeze2018}, and current-density imaging \cite{Pollok2025,Zhong2024,Chatzidrosos2019}. High-density NV ensembles are also attractive as macroscopic spin sensors for searching for exotic interactions and beyond-standard-model physics \cite{Liang2023,Guo2024,Cong2025,Chigusa2025,JacksonKimball2020,Astner2018}.

This increase in spin density, however, comes with an important tradeoff. At high defect concentrations, the ensemble dynamics are strongly affected by homonuclear dipolar interactions among NV centers\cite{Barry2020,Hughes2025,Zhao2024R,Zhao2023,Zhao2020,Biter2025} and by heteronuclear-like coupling to the surrounding spin bath, including P1 centers and residual nuclear spins \cite{Knowles2014,Cooper2020,Rosenfeld2018}. Together with static disorder from strain, local electric fields, and magnetic impurities, these interactions lead to rapid inhomogeneous dephasing and limit the usable coherence time of the sensor \cite{Kucsko2018,Zhao2012}. As a result, increasing the number of sensing spins does not automatically improve sensitivity. The gain from larger signal size can be offset by reduced coherence. 

A standard strategy to extend spin coherence is dynamical decoupling \cite{Choi2020,Rizzato2023,Yao2026,McGuire2025,Mitrikas2014}. However, conventional echo-based sequences such as Carr-Purcell-Meiboom-Gill (CPMG) or XY-type protocols \cite{Carr1954,Meiboom1958,Gullion1990} are not well suited for direct dc magnetometry, because their $\pi$-pulse sign inversions refocus the phase from both quasistatic noise and the target dc field. Thus, they can protect coherence while simultaneously canceling the signal of interest. Dense-ensemble dc sensing therefore requires a control protocol that suppresses spin-spin interactions without eliminating the response to a static detuning.

For a CPMG or XY-type cycle, the toggling-frame sensing operator averages to zero, $\bar{S}_z^{(0)}=0$, so that the target dc term is refocused, while the secular homonuclear dipolar Hamiltonian remains unchanged, $\bar{H}_{dd}^{(0)}=H_{dd}$. In contrast, ideal WAHUHA gives $\bar{S}_z^{(0)} = \frac{1}{3} (S_x + S_y + S_z) , \bar{H}_{dd}^{(0)} = 0$, thereby retaining a dc response reduced by 1/$\sqrt{3}$ while canceling the homonuclear dipolar interaction to zeroth order.

The Waugh-Huber-Haeberlen (WAHUHA) sequence provides such a protocol. Originally developed in solid-state NMR to average homonuclear dipolar interactions \cite{Waugh1968}, WAHUHA cyclically rotates the spin quantization axis through different spatial directions rather than simply alternating between $+z$ and $-z$. In the ideal-pulse limit, this symmetrizes the homonuclear dipolar Hamiltonian while transforming a static detuning term along $z$ into a nonzero effective field in the rotating frame. For a dc field coupled to $S_z$, the ideal WAHUHA average transforms the sensing term into an effective sensing field proportional to  $S_x + S_y + S_z$. The resulting dc response is reduced by a factor of $1/\sqrt{3}$ relative to the original $S_z$ signal, but remains finite along the [111] direction (Refer to Appendix \ref{app:AHT} for a detailed derivation). Thus, unlike conventional echo-based decoupling, WAHUHA can in principle suppress interaction while preserving a dc sensing channel \cite{Balasubramanian2019,Oon2026}. 
 
More broadly, WAHUHA-type control can be viewed as an instance of Floquet Hamiltonian engineering, in which a periodically driven system is designed to evolve under an effective Hamiltonian that differs from its static Hamiltonian \cite{Shirley1965,Goldman2014,Bukov2015,Kim2026,Rovny2018,Zhang2017,Jotzu2014l,Duan2025}. In the ideal-pulse limit, average Hamiltonian theory (AHT) provides a simple and powerful description of this process and predicts the leading-order cancellation or symmetrization of unwanted interactions \cite{Haeberlen1968}. This ideal picture motivates the use of WAHUHA for dense NV-ensemble dc sensing \cite{Balasubramanian2019,Oon2026}. The sequence can suppress the homonuclear interaction while preserving a finite detuning-to-phase response.

Realistic NV-ensemble experiments implementations of WAHUHA, however, operate outside this idealized limit. The microwave pulses have finite duration and the spin state is sampled stroboscopically only after each control cycle. The measurement therefore tracks the dynamics only at integer multiples of the driving period: it does not directly resolve the high-frequency intra-cycle micromotion but instead captures its net contribution to the cycle-to-cycle evolution. This makes the phase associated with the effective Floquet Hamiltonian the relevant quantity for the observed stroboscopic response \cite{Shirley1965,Bukov2015,Geier2021,Schweizer2019,Flaschner2016,Oka2019,Duan2025}. During both the free-evolution windows and the finite-width pulses, the spins accumulate phase from detuning, hyperfine shifts, residual interactions, and control imperfections. Consequently, the measured stroboscopic signal is governed by the full on-cycle unitary rather than by the zeroth-order average Hamiltonian alone. This distinction becomes especially important when the decay envelope is used as a proxy for sensing performance: an extended effective dephasing time under periodic driving does not necessarily imply enhanced dc magnetic-field sensitivity.

Here we investigate this distinction in a dense diamond NV ensemble controlled by the WAHUHA sequence. Experimentally, WAHUHA extends the observed stroboscopic decay time from the Ramsey inhomogeneous dephasing time, $T_{2}^* \approx 0.9\ \mu$s, to an effective inhomogeneous dephasing time, $T_{2,eff}^* \approx 31\ \mu$s. Despite this more than thirty-fold increase in the effective dephasing time, the corresponding dc magnetic-field sensitivity shows little improvement. To understand this discrepancy, we perform detuning-resolved stroboscopic spectroscopy and analyze the measured response using a finite-pulse Floquet model. By constructing the one-cycle Floquet unitary, we map the detuning-dependent eigenphase, $\Phi(\Delta,\tau)$, and show that the long-lived signal arises from phase wrapping and quasienergy branch folding in the stroboscopic readout. These effects reshape the spectrum while suppressing the detuning-to-phase transduction slope, $d\Phi/d\Delta$, which governs the dc magnetic-field response. Our work, therefore, establishes finite-pulse Floquet analysis as a practical framework for distinguishing extended effective dephasing from metrological gain in periodically driven spin ensembles. 

\section{Floquet description of stroboscopic coherence and sensitivity}\label{sec2}
\begin{figure*}
\centering
\includegraphics[width=0.9\textwidth]{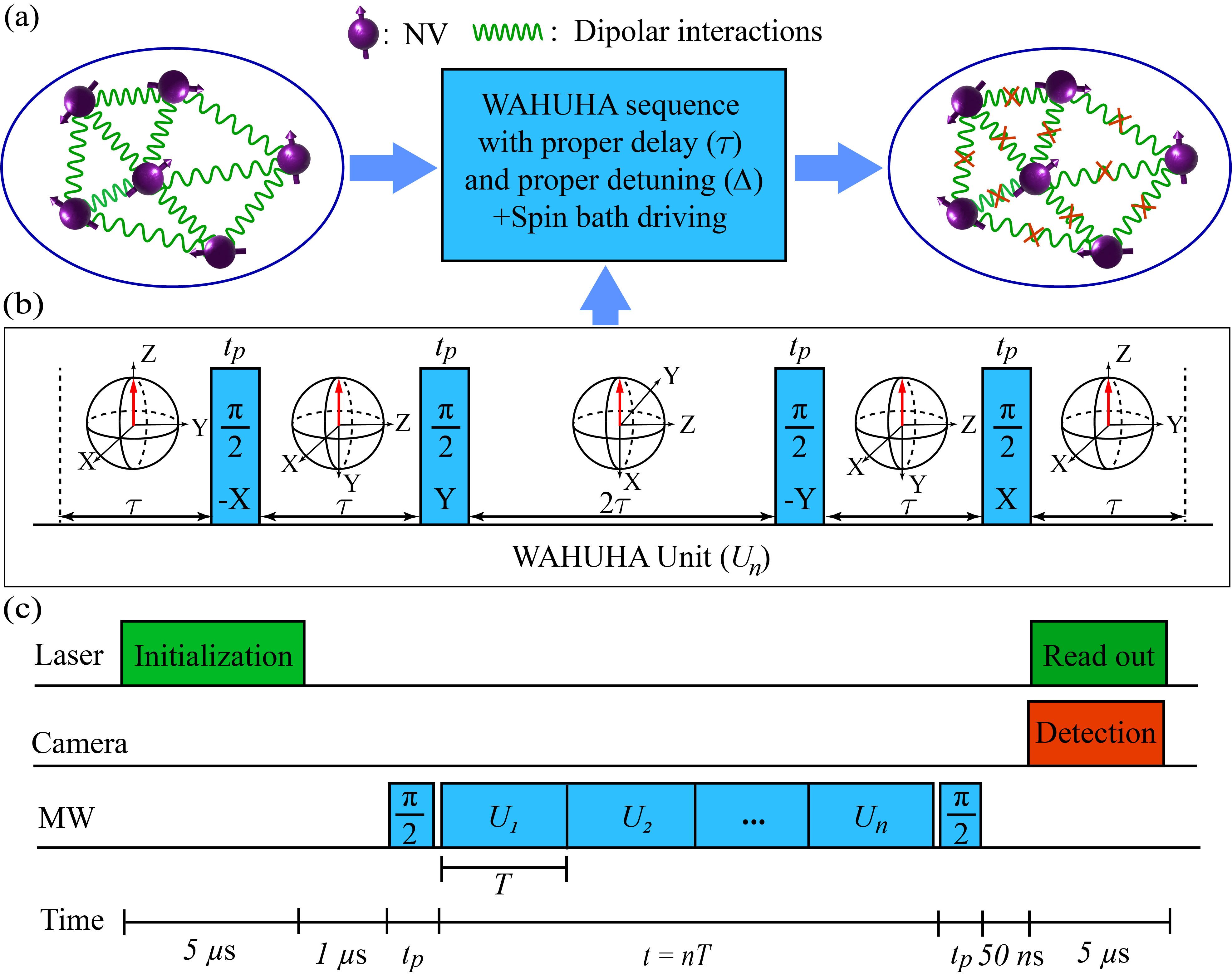}
\caption{\label{fig1}WAHUHA control and stroboscopic measurement protocol in a dense NV ensemble. (a) Conceptual schematic of a dipolar-coupled NV ensemble under periodic WAHUHA control. The sequence symmetrizes the homonuclear dipolar interaction while in the experimentally relevant finite pulse regime, the observed response is more naturally described in terms of the one-cycle Floquet unitary. Additional spin-bath driving can suppress the residual contribution from P1 spins. (b) Pulse timing and phase configuration of a single WAHUHA unit, consisting of four $\pi/2$ microwave pulses with phases -X, Y, -Y, and X, separated by delays $\tau$, $\tau$, $2\tau$, $\tau$, and $\tau$. Therefore, one-cycle duration becomes $T = 6\tau+4t_p$, where $t_p = 10$ ns is the $\tau/2$-pulse duration. The Bloch-sphere sketches illustrate the cyclic trajectory of the effective spin under one control period. (c) Experimental timing diagram. After optical initialization and an initial $\pi/2$ pulse, the ensemble evolves under $n$ WAHUHA cycles for a total time $t=nT$. A final $\pi/2$ pulse projects the stroboscopic state for optical readout, and the resulting fluorescence defines the stroboscopic data point $S[n]$.}
\end{figure*}

The experimentally relevant observable in our WAHUHA protocol is the stroboscopic response measured after each cycle of the periodic pulse sequence (Fig. \ref{fig1}). In the ideal limit, WAHUHA averages the homonuclear NV-NV dipolar interaction to leading order, while auxiliary spin-bath driving suppresses the contribution from P1 spins \cite{Bauch2018,DeLange2012}. Average Hamiltonian theory (AHT) provides a useful leading-order description in the fast-driving regime. More specifically, the conventional zeroth-order WAHUHA approximation is reliable when the interpulse delay is short compared with the characteristic disorder- and interaction-induced evolution timescales, $\tau \ll W^{-1},J^{-1}$, where $W$ and $J$ denote on-site disorder and spin-spin interaction, respectively \cite{Zhao2020}. As $\tau$ increases, higher-order AHT approximation loses quantitative accuracy. We therefore describe the experimentally relevant dynamics using the finite-pulse Floquet unitary. Unlike the zeroth-order AHT approximation, the direct evaluation of the one-cycle unitary does not require a short-$\tau$ expansion and can be applied at arbitrary interpulse delay within the effective two-level Hamiltonian in Eq. (\ref{eq1}). The respective ranges of applicability of the two descriptions are discussed in Appendix \ref{app:ValidityAHT}. The principal symbols used below are summarized in Table \ref{tab1} of Appendix \ref{sec:AppA}.

We model the driven NV ensemble within the $m_s = 0$ and $m_s =-1$ manifold as an effective two-level system. In the rotating frame, the control Hamiltonian during the pulse sequence is written as 
\begin{equation}
H_{pulse} = \frac{\Delta}{2} \sigma_z + \frac{\Omega}{2} \left( \cos(\phi) \sigma_x + \sin(\phi) \sigma_y \right) + \frac{A_{zz}}{2} \sigma_z I_z,
\label{eq1}
\end{equation}
where $\Delta$ is the effective microwave detuning, $\Omega$ is the Rabi frequency, $\phi$ is the pulse phase and $A_{zz} I_{z}$ is the hyperfine interaction with the nitrogen nuclear spin. Because the pulse sequence is periodic, with a duration of one-cycle $T = 6\tau+4t_{p}$, the relevant dynamics are characterized by the propagator of one-cycle.
\begin{equation}
U(T)\equiv\mathcal{T}\exp\left[-i\int_0^T H_{pulse} \left(t\right)dt\right]=\exp{\left(-iH_F T\right)},
\label{eq2}
\end{equation}
which defines the Floquet Hamiltonian. In the numerical calculation, Eq. (\ref{eq2}) is evaluated directly as the ordered production of the free-evolution and finite-pulse segment propagators. It therefore retains the detuning, hyperfine shift, and finite-duration pulse evolution contained in $H_{pulse}(t)$ without truncating a Magnus expansion. In the short-delay regime, the resulting Floquet phase is consistent with the leading-order AHT prediction. At longer delays, however, the exact one-cycle phase can substantially deviate from the linear, unwrapped AHT picture and can extend beyond the principal phase interval ($-\pi,\pi$]. 

For an effective two-level system, $H_F=\frac{1}{2}\mathbf{w}\cdot\boldsymbol{\sigma}$, so that the experimentally relevant quantity is the Floquet eigenphase splitting, $\Phi(\Delta,\tau) = |\mathbf{w}(\Delta,\tau)|T$, which determines the stroboscopic spectral response measured in the experiment.  

To extract the relevant Floquet phase, we evaluate the eigenphase splitting of the one-cycle propagator. For a 2×2 unitary matrix $U(T)$, the overall global phase is removed by defining $V = U/ \sqrt{\operatorname{det}(U)}$. The resulting unitary matrix can be written as 
\begin{equation}
 V=e^{-i \frac{\Phi}{2} \mathbf{\hat{n}} \cdot \boldsymbol{\sigma}} = \cos\left(\frac{\Phi}{2}\right)\mathbf{I}-i\sin\left(\frac{\Phi}{2}\right)\ \hat{\mathbf{n}} \cdot\boldsymbol{\sigma},
\label{eq3}
\end{equation}
where $\hat{\mathbf{n}} = \mathbf{w}/|\mathbf{w}|$.
The Floquet eigenphase splitting, $\Phi$, enables a direct evaluation of the quasi-energy phase for arbitrary finite pulse width, imperfections, and detuning errors, and forms the basis of the detuning-resolved Floquet spectroscopy presented below. For a more detailed derivation and discussion of the Floquet eigenphase splitting, see Appendix \ref{app:FloquetUnitary}.

The measured signal is sampled stroboscopically after each WAHUHA cycle. We model the response after $n$ cycles as 
\begin{equation}
S[n]=S_{off} + C(\tau) e^{-t/T_{2,eff}^*} \cos\left[n  \Phi_{wr} (\Delta,\tau)+\phi_r\right]  ,
\label{eq4}
\end{equation}
where $C(\tau)$ is the operational constant, $T_{2,eff}^*$ is the effective inhomogeneous dephasing time, $\phi_r$ is a readout-phase offset, and $ \Phi_{wr} \in (-\pi,\pi]$ is the wrapped Floquet phase relevant to stroboscopic sampling. The corresponding Fourier peak is located at 
\begin{equation}
f_p\ (\Delta,\tau) =  \frac{\Phi_{wr}(\Delta,\tau)}{2\pi T},
\label{eq5}
\end{equation}
up to the sign convention of the Fourier transform. In the small-detuning regime, this allows the detuning-dependent spectral branch to be parametrized directly in terms of the Floquet phase landscape.

In the weak-phase-accumulation regime, the unwrapped Floquet phase can be expanded as 
\begin{equation}
\Phi(\Delta,\tau) \approx \Phi_0 + 2\pi \kappa(\tau) \Delta T, 
\label{eq6}
\end{equation}
where $\Phi_0$ denotes a cycle-phase intercept and $\kappa(\tau)$ characterizes the detuning-to-phase conversion. In this limit, the dominant spectral branch is approximately linear in detuning, and the corresponding zero-frequency crossing occurs at
\begin{equation}
\Delta_{cross} (\tau)=\frac{-\Phi_0}{2\pi\kappa T }.  
\label{eq7}
\end{equation}

This expression immediately explains why the detuning-axis offset if most pronounced at short interpulse delays. For a finite phase intercept, the crossing detuning scales inversely with the cycle period $T$. As $\tau$ increases and $T$ becomes longer, the same phase intercept produces a progressively smaller detuning shift, causing the spectral branch to approach the $\Delta = 0$ axis.

\begin{figure*}
\centering
\includegraphics[width=0.85\textwidth]{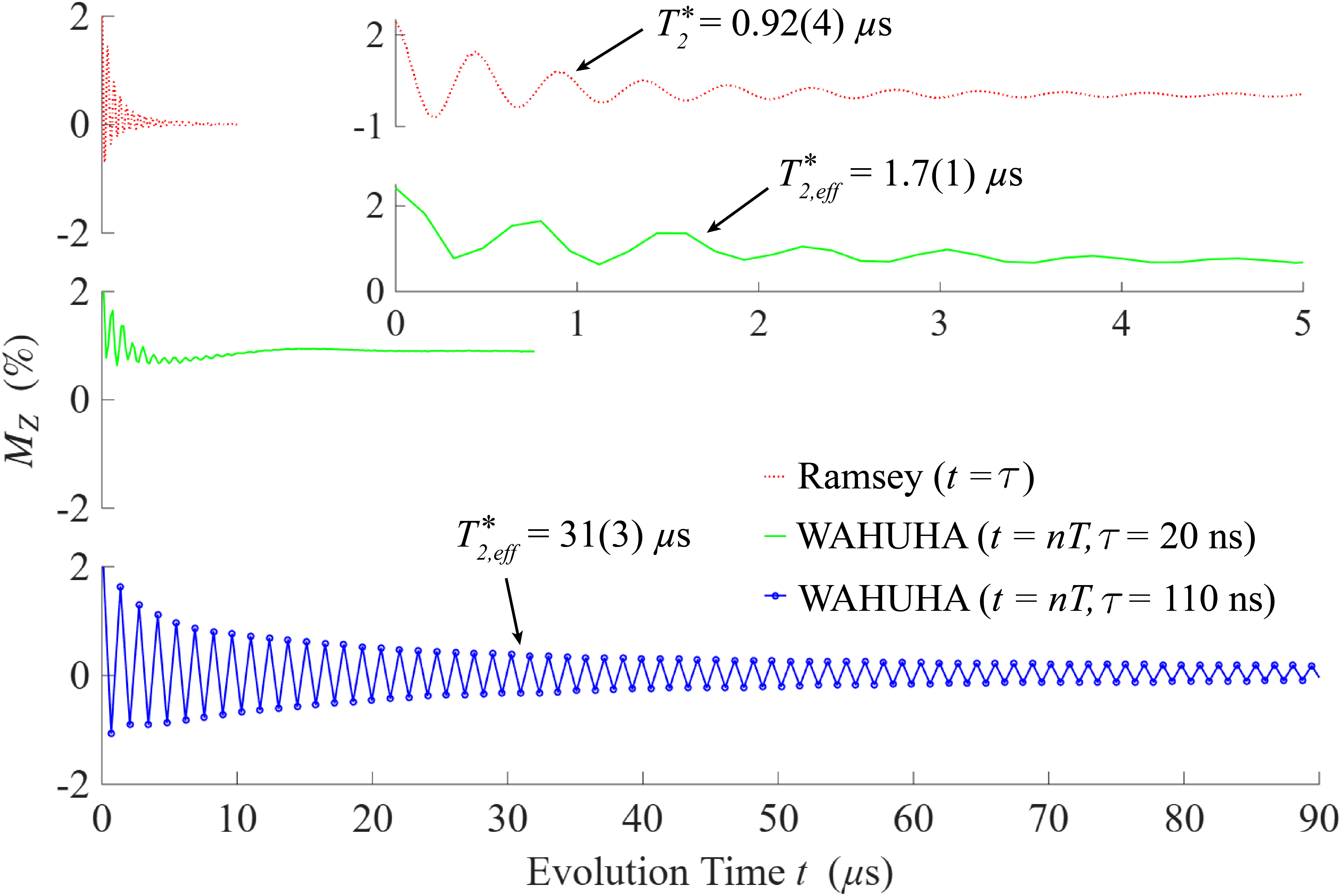}
\caption{\label{fig2}Effective inhomogeneous dephasing under WAHUHA. The longitudinal magnetization $M_Z$ is measured as a function of evolution time for Ramsey and WAHUHA protocols. The Ramsey free-induction signal yields an intrinsic inhomogeneous dephasing time $T_2^\ast= 0.92(4)\ \mu$s (top, red). Under WAHUHA control, the fitted stroboscopic decay constant increases to  $T_{2,eff}^\ast = 1.7(1)\ \mu$s for $\tau = 20$ ns (middle, green) and  $T_{2,eff}^\ast =  31(3)\ \mu$s for $\tau$ = 110 ns (bottom, blue). The inset compares the first 5 $\mu$s  of the Ramsey and short-delay WAHUHA traces. Solid curves are nonlinear least-squares fits to $S(t)=A\exp{\left[-\left({t}/{T_{dec}}\right)\right]}\cos{\left(2\pi ft+\phi_{f}\right)}+C $. The sample single-exponential envelope was used for all traces. $T_{dec}$ is denoted $T_{2}^\ast$ for Ramsey and $T_{2,eff}^\ast$ for WAHUHA. The $T_{2,eff}^\ast$ of 31 $\mu$s value demonstrates long-lived stroboscopic envelope produced under WAHUHA control.}
\end{figure*}

The decreasing accuracy of the zeroth-order AHT approximation and the appearance of quasienergy folding are distinct effects. The former originates from higher-order and finite-pulse contributions that are neglected in the leading-order approximations, whereas the latter follows from the modulo-2$\pi$ structure of the exact Floquet eigenphases. Both effects become relevant as the cycle period increases, but quasienergy folding is not itself the cause of the AHT truncation error.

As the interpulse delay is further increased, the accumulated phase over one-cycle is no longer small, and the observed dynamics are governed by the wrapped phase, $\Phi_{wr}$. This phase wrapping is a direct consequence of the Nyquist sampling theorem, which limits the bandwidth of stroboscopic measurements. When the unwrapped phase reaches the boundary $\Phi=(2m+1)\pi$, the corresponding spectral branch is folded back into the finite stroboscopic frequency window, producing the branch wrapping and folding observed in the detuning-resolved spectra. In particular, near the $\pi$-phase boundary, we may write $\Phi(\Delta,\tau)=\pi + \delta \Phi(\Delta,\tau)$ such that the oscillatory component of the signal becomes $\cos(n\Phi+\phi_r)=(-1)^n \cos(n \delta\Phi+\phi_r)$. The factor $(-1)^n$ gives rise to the characteristic $2T$ stroboscopic response, while the residual phase ($\delta\Phi$) determines the slow evolution of the envelope within that period-doubled frame.

This structure has an immediate consequence for metrology. Differentiating Eq.~(\ref{eq4}). with respect to detuning yields
\begin{equation}
\frac{\partial{S[n]}}{\partial{\Delta}}=-C(\tau) e^{-nT/T_{2,eff}^*} \sin \left[{n\Phi_{wr}(\Delta,\tau)+\phi}\right] n\frac{d\Phi_{wr}}{d\Delta},   
\label{eq8}
\end{equation}

At the optimal operating point, the dc transduction is therefore controlled not by the $T_{2,eff}^\ast$, but by the product of contrast, $T_{2,eff}^\ast$, and Floquet transduction slope,  $\frac{d\Phi_{wr}}{d\Delta}$. Near the $\pi$-phase boundary, where $\Phi=\pi+\delta\Phi$, the signal becomes locally quadratic in $\delta\Phi$, so that the linear response to detuning is strongly suppressed even when the stroboscopic oscillations remain long lived. This distinction between an effective dephasing time and a reduced detuning-to-phase transduction slope is central to the interpretation of our measurements below. These considerations provide a Floquet framework for interpreting the detuning-resolved stroboscopic spectra, the emergence of the $2T$ response, and the discrepancy between the extended $T_{2,eff}^\ast$ and the actual dc-magnetic field sensitivity.

\section{Results and discussion}\label{sec3}

\subsection{\label{subsec:effDephasing}Effective inhomogeneous dephasing under WAHUHA control}

\begin{figure*}
\centering
\includegraphics[width=0.9\textwidth]{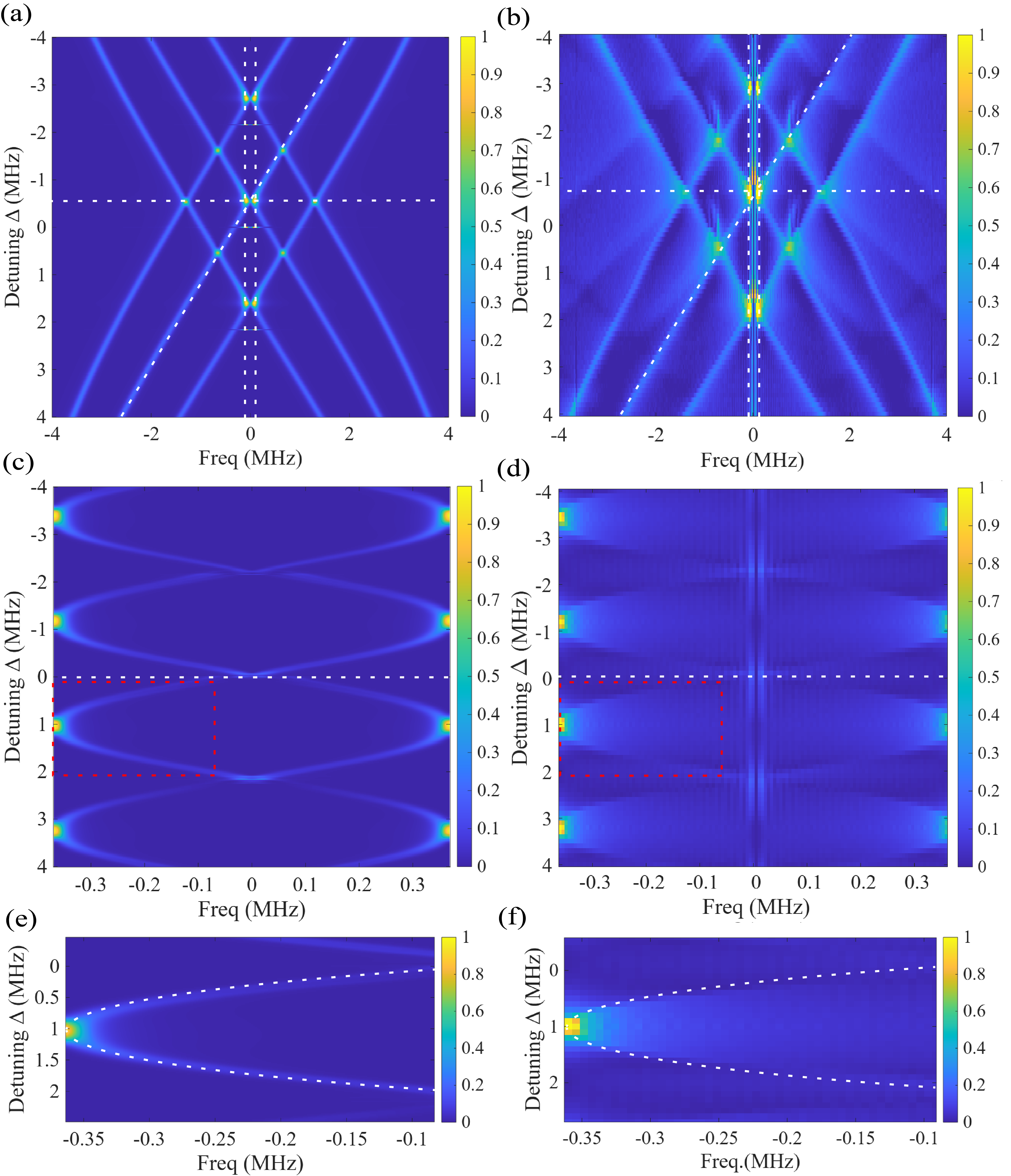}
\caption{\label{fig3}Detuning-resolved stroboscopic spectra under the WAHUHA sequence. The spectra are obtained by Fourier transforming the measured stroboscopic magnetization,$M_Z (nT)$ with respect to the Floquet cycle number $n$, where the WAHUHA cycle period is defined as  $T = 6\tau+4t_p$ with a pulse duration of $t_p$ = 10 ns (equivalent to $\Omega_{\text{Rabi}} = 25\text{ MHz}$). (a, b) Simulated and experimental spectra for a short interpulse delay of of  $\tau$ = 12 ns. In this regime, the dominant spectral branches are approximately linear in detuning and follow the expected short-delay scaling, $f_{p}\approx \Delta /\sqrt{3}$. The white dashed lines explicitly indicate the primary Floquet branch and the horizontal zero-detuning axis ($\Delta = 0$). (c, d) Simulated and experimental spectra for a longer interpulse delay of $\tau$ = 220 ns, shown within the narrower Nyquist frequency window, [-1/(2$T$), 1/(2$T$)]. In this regime, larger phase accumulation per cycle drives the Floquet branch toward the $\pi$-phase boundary, leading to phase wrapping and branch folding into the narrower Nyquist window, which appear as periodic oval-shaped spectral structures. (e,f) Enlarged views of the regions marked in (c,d) near a Nyquist-edge region. The zoomed panels resolve the curved finite-pulse Floquet dispersion used to evaluate the local detuning-to-phase transduction. The agreement between simulation and experiment supports a Floquet interpretation of the observed stroboscopic response.}
\end{figure*}

We compare the periodically driven NV ensemble with a conventional Ramsey free-induction measurement. To quantitatively extract the Ramsey dephasing time $T_2^\ast$  and the effective stroboscopic decay time $T_{2,eff}^\ast$, all magnetization traces in Fig. \ref{fig2} were fitted using the same damped single-exponential model, $S(t)=A\exp{\left[-\left(\dfrac{t}{T_{dec}}\right)\right]}\cos{\left(2\pi ft+\phi_{f}\right)}+C $, where  $T_{dec} = T_2^\ast$ for the Ramsey measurements, $T_{dec} = T_{2,eff}^\ast$ for the WAHUHA measurements. For the WAHUHA data, $t = nT$, where $n$ is the number of WAHUHA cycles and $T = 6\tau+4t_p$. Here, $A$ is the amplitude, $C$ is a baseline offset, $f$ is the oscillation frequency, and $\phi_{f}$ is the phase offset. As shown in Fig. \ref{fig2} (red line), the Ramsey signal yields an intrinsic inhomogeneous dephasing time of 0.92 $\mu$s. This is consistent with the strong static disorder and dipolar broadening expected in the dense NV ensemble, $[\text{P1}_{ini}]$ = 12 ppm and [NV] = 3.6 ppm. But, under repeated WAHUHA protocol the stroboscopic signal develops a longer-lived envelope. For a short interpulse delay of 20 ns, the fitted decay constant increases to $T_{2,eff}^\ast \approx 1.7\ \mu$s, as shown in Fig. \ref{fig2} (green line). This scale is broadly consistent with a partial suppression of the homonuclear contribution while leaving residual dephasing from the environment. 

In contrast, when the interpulse delay is increased to 110 ns, the fitted decay constant abruptly increases to $T_{2,eff}^\ast\ \approx 31\ \mu$s, as shown in Fig. \ref{fig2} (blue line), far beyond the timescale expected from suppression of the NV-NV interaction alone. Here, $T_{2,eff}^\ast$ is defined as a phenomenological decay constant obtained from the stroboscopically sampled WAHUHA envelope.  It is not equivalent to the Ramsey inhomogeneous dephasing time $T_2^\ast$, because the stroboscopic envelope depends not only on physical decoherence but also on the detuning dependence of the one-cycle Floquet phase and the particular Floquet branch sampled by the stroboscopic readout. In the long-delay regime, the large fitted value is associated with the folded structure of Floquet quasienergy branches. In the same regime, the detuning-to-phase transduction slope, $d\Phi/d\Delta$, is strongly reduced. Accordingly,  $T_{2,eff}^\ast$ characterize the persistence of the driven, sampled response rather than an intrinsic extension of the free-induction coherence time.

To identify the mechanism underlying the long-lived stroboscopic signal, we map the Fourier spectrum of the WAHUHA response as a function of microwave detuning, $\Delta$. For each detuning, the magnetization, $M_Z (nT)$, is measured as a function of Floquet cycle number, $n$, and Fourier transformed along the stroboscopic index, where $T = 6\tau + 4 t_p$ is the cycle period. Representative spectra are shown in Fig. \Ref{fig3} for short and long interpulse delays. In the short-$\tau$ regime, $\tau$ = 12 ns, the spectra are dominated by nearly linear Floquet branches. In this limit, the one-cycle phase remains small and depends approximately linearly on detuning, Eq. (\Ref{eq6}), so that the dominant branch follows the expected short-delay scaling, $f_{p}\approx\Delta/\sqrt3$. The white dashed lines highlight this primary Floquet branch as well as the horizontal zero-detuning axis ($\Delta=0$).

The behavior changes qualitatively at longer interpulse delay, $\tau$ = 220 ns. As the free evolution windows becomes longer, the phase accumulated within a single cycle grows much more rapidly with detuning. As a result, the spectral branch is compressed into the narrower Nyquist window imposed by the stroboscopic sampling rate, and once the accumulated phase approaches the $\pi$-phase boundary, it is folded back into the observable frequency range. This produces the periodic, oval-shaped spectral structures observed in both experiment and finite-pulse simulation. In this regime, the signal is naturally described by a wrapped Floquet phase ($\Phi_{wr}\in(-\pi,\ \pi]$), $S[n] \propto \cos{(n\Phi_{wr}(\Delta,\tau)+\phi_r)}$, and near the boundary $\Phi=\pi+\delta\Phi$, takes the form $S[n] \propto(-1)n \cos(n\delta\Phi+\phi_r)$, revealing the origin of the observed 2$T$ response. The appearance of a 2$T$ signal might resemble subharmonic dynamics associated with discrete time crystals \cite{Yao2017,Zhang2017}. However, in our experiment the period-doubled response in our experiment does not evidence of a many-body time-crystalline phase. Instead, it arises from phase wrapping and branch folding of the single-cycle Floquet quasienergy spectrum. In this context, the 2$T$ response is a stroboscopic spectral effect of the driven sensor, rather than a rigid interaction-stabilized subharmonic phase. Consequently, these spectra demonstrate that the long $T_{2,eff}^*$\ observed at larger delays is tied to the quasienergy branch structure of the driven system.

The short-delay spectra also exhibit two robust features that are absent from the idealized linear picture, a finite detuning-axis offset and a small avoided-crossing-like gap near the branch center. Both are reproduced by a minimal finite-pulse Floquet description in which the one-cycle phase contains a residual longitudinal intercept, $\Phi_0$, together with a transverse error scale, $\Phi_{gap}$. In this picture, the longitudinal intercept shifts the zero-frequency crossing away from $\Delta=0$, while $\Phi_{gap}$ pins the minimum splitting at the center of the branch. As $\tau$ increases, the crossing detuning approximately follows $\Delta_{cross} \propto -1/T$  consistent with the experimentally observed recovery of the branch center toward zero detuning, Fig. \ref{fig4} (c). These measurements show that the dramatic increase in $T_{2,eff}^\ast$ arises from phase wrapping and quasi-energy branch folding in the stroboscopic readout.

\subsection{\label{subsec:FloquetBranch}The avoided crossing and zero-frequency crossing of the Floquet branch}
\begin{figure*}
\centering
\includegraphics[width=0.9\textwidth]{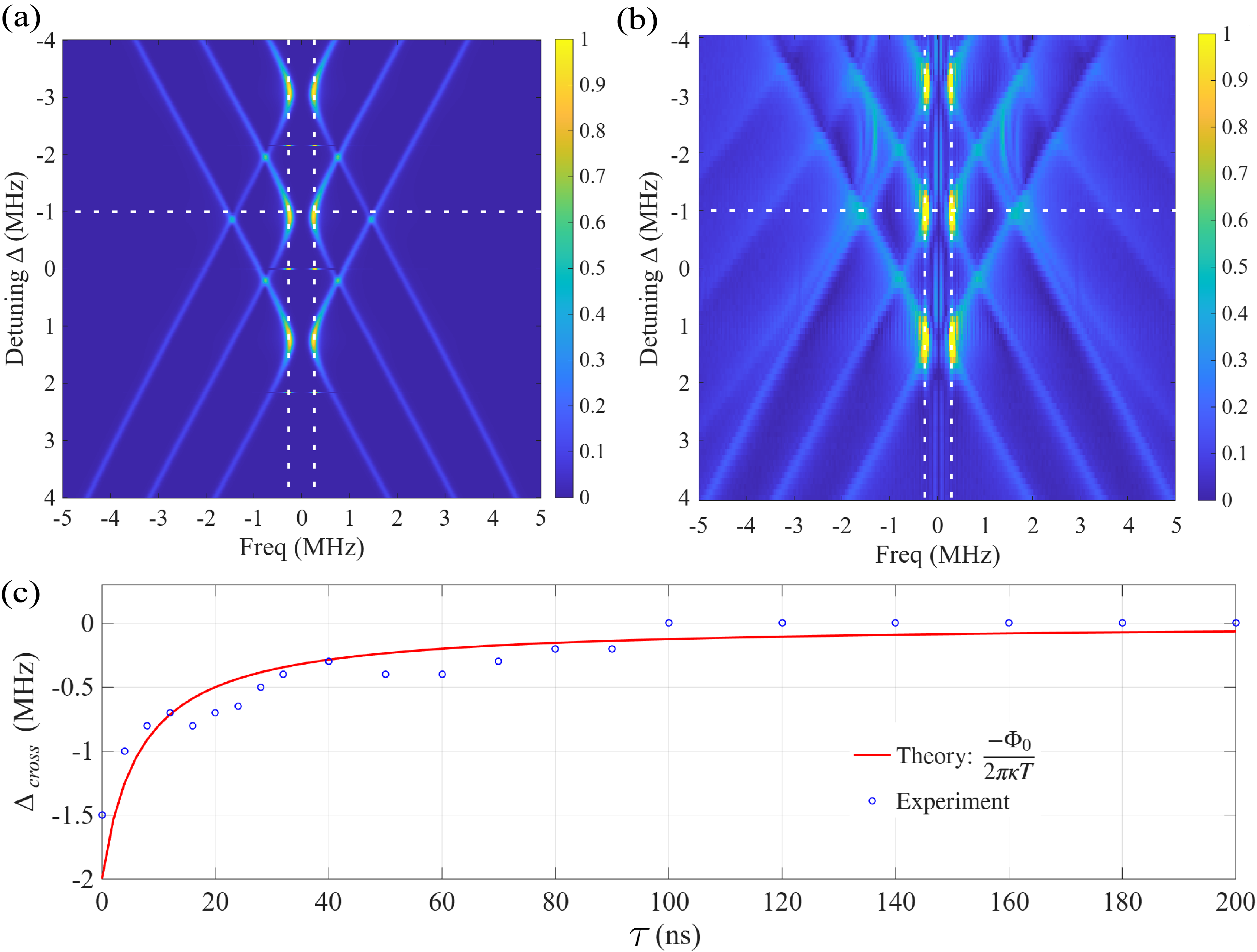}
\caption{\label{fig4}Short-delay Floquet branch structure and zero-frequency crossing. (a) Simulated and (b) experimental FFT spectra of the $M_Z$ under the WAHUHA sequence for a short interpulse delay of $\tau$ = 4 ns. The spectra display the detuning dependent Floquet branch as a function of microwave off-resonance detuning ($\Delta$). Dashed lines indicate the zero-detuning axis and the central branch region, highlighting the finite gap and the shifted branch center. (c) Detuning value ($\Delta_{cross}$) at which the dominant branch crosses zero frequency ($f = 0$), plotted as a function of $\tau$. Experimental data (blue circles) are compared with the analytical prediction (red line, $-\mathrm{\Phi}_0/2\pi\kappa T$).}
\end{figure*}
By mapping the effective nutation frequencies in the stroboscopic Fourier spectrum, we resolve distinct branches that track the detuning-dependent Floquet response of the driven NV ensemble. In the short-delay regime, the measured branch cannot be described by a linear phase accumulation alone. Instead, the data is captured by a minimal finite-pulse Floquet phenomenology in which the one-cycle phase contains both a residual longitudinal intercept and a transverse error scale,
\begin{equation}
\Phi(\Delta,\tau)=\sqrt{(\Phi_0+2\pi\kappa\Delta\ T)^2+\Phi_{gap}^2 },   
\label{eq9}
\end{equation}
where $\Phi_0$ is a cycle-phase offset, $\kappa$ is the detuning-to-phase conversion factor, $T$ is the WAHUHA cycle period, and $\Phi_{gap}$ parameterizes the finite minimum splitting of the branch arising from pulse imperfections. Specifically, at high Rabi frequencies, the control pulses deviate from an ideal square profile due to transient distortions, causing the ringing of a preceding pulse to interfere with the subsequent one. In this description, $\Phi_{gap}$ produces the observed avoided-crossing-like gap, whereas $\Phi_0$ shifts the symmetry axis of the branch away from $\Delta=0$.

In the idealized linear limit $\Phi_{gap}=0$, the accumulated phase would pass continuously through zero, giving rise to the sharp intersecting branches seen in the short $\tau$ simulation. In the experiment, however, the finite transverse term prevents the phase from vanishing at the branch center and instead pins the minimum splitting at $\left|\Phi_{gap}\right|$, thereby producing the small gap in Fig. \ref{fig4} (a, b) due to control imperfections. The detuning value at which the longitudinal contribution vanishes is obtained from $\Phi_0+2\pi \kappa \Delta\ T=0$,which yields the zero-frequency crossing $\Delta_{cross}(\tau)=-\Phi_0 / 2\pi \kappa T$, Eq. (\ref{eq7}).

Fig. \ref{fig4} (c) shows that this simple scaling captures the dominant trend of the data. At short interpulse delays, the small cycle period T makes $\Delta_{cross}$ sensitive to the cycle-phase intercept, resulting in a pronounced shift away from zero detuning. As $\tau$ increases, the same phase intercept produces a smaller detuning offset, and $\Delta_{cross}$ to approaches zero. The gap and the detuning-axis offset provide an experimental signature of finite pulse Floquet structure in the short-delay regime. 

\subsection{\label{subsec:DCSensitivity}Effective coherence and dc sensitivity}
\begin{figure*}
\includegraphics[width=0.9\textwidth]{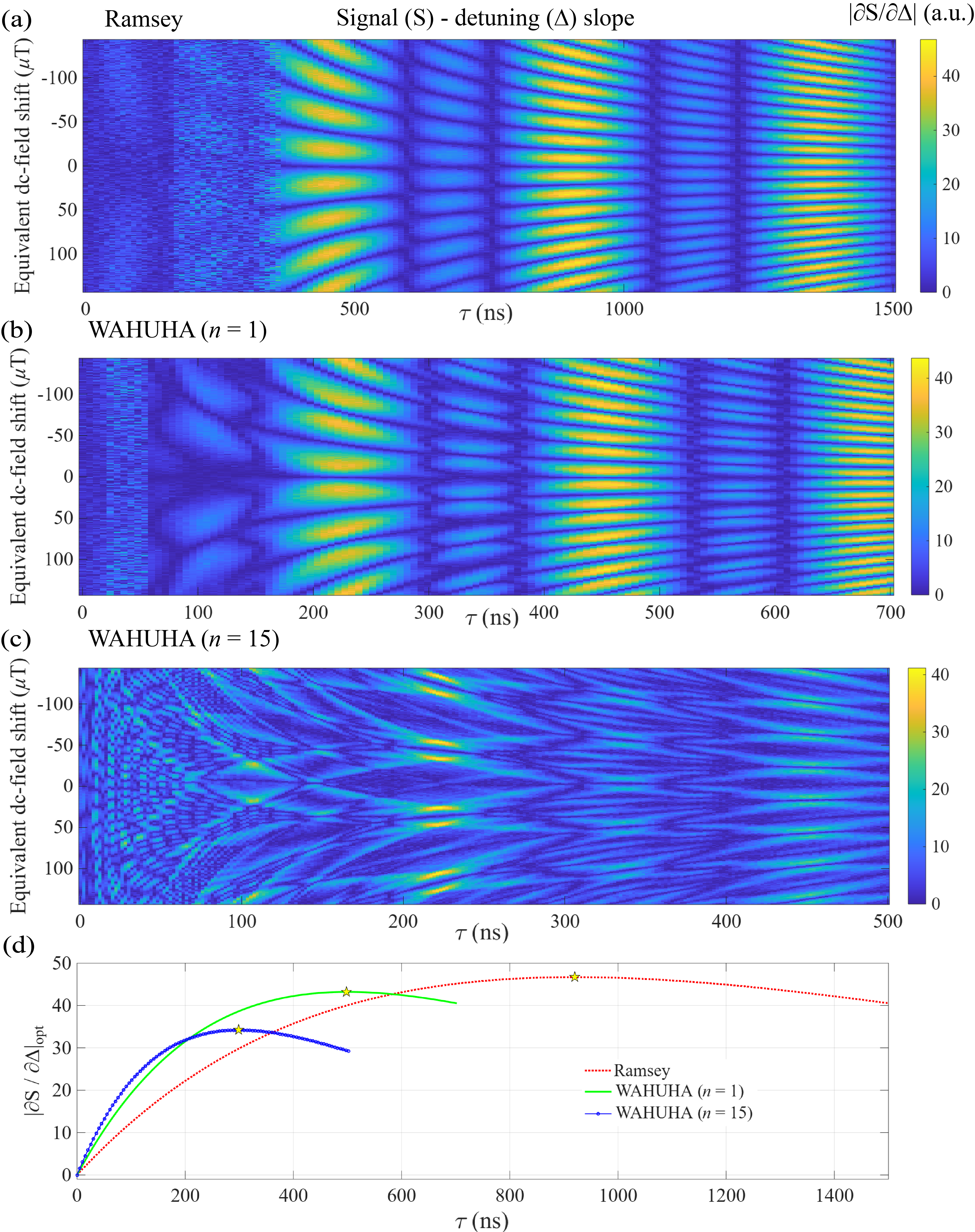}
\centering
\caption{\label{fig5}Experimentally measured detuning-response slope for the Ramsey and WAHUHA protocols. Two-dimensional maps of the signal slope are plotted as a function of interpulse delay ($\tau$) and microwave detuning ($\Delta$). The detuning $\Delta$ serves as a calibrated proxy for an external dc magnetic field. (a) Standard Ramsey sequence. (b) Single WAHUHA unit ($n = 1$). (c) Multi-pulse WAHUHA sequence($n=15$). In each panel (a, b, c), the vertical line marks the optimal $\tau$ at which the maximum slope is obtained. (d) Signal derivative envelopes $|\partial S/\partial\Delta|_{opt}$ as a function of pulse sequence duration $\tau$ for Ramsey (red dotted line), WAHUHA ($n$ = 1, green solid line), and WAHUHA ($n$ = 15, blue open circles) sequences.  Although WAHUHA strongly increases the effective dephasing time, $T_{2,eff}^*$, the maximum slope does not increase correspondingly. Instead, the long $\tau$ regime exhibits a fragmented response landscape associated with Floquet phase wrapping and  suppression of the detuning-to-phase transduction slope, $d\mathrm{\Phi}/d\Delta$. These maps explains why the extended $T_{2,eff}^*$ does not translate into enhanced dc sensitivity.}
\end{figure*}
The magnetic field sensitivity, $\eta_B$, can be written as below, 
\begin{equation}
\eta_B\equiv\delta B_{min}\sqrt{t_m}=\frac{\sigma_S}{{\partial S}/{\partial B}}\sqrt{t_m} = \frac{\sigma_S}{(\partial S/\partial\Delta)(\partial\Delta/\partial B)}\sqrt{t_m}, 
\label{eq10}
\end{equation}
where $\delta B_{min}$ is the minimum detectable magnetic field, $t_m$ is the measurement time, and $\sigma_S$ is the total signal noise, including photon shot noise and electronic noises. From Eq. (\ref{eq8}), the dc transduction is controlled not by the envelope lifetime alone, but by the product of the operational contrast, the envelope amplitude, and the Floquet transduction slope $d\Phi/d\Delta$. At the optimal readout, the maximum response scales as
\begin{equation}
\left. \frac{\partial S[n]}{\partial \Delta} \right|_{max}
=-C(\tau) e^{-t/T_{2,eff}^*} \frac{t}{T} \left|\frac{d\Phi_{wr}}{d\Delta}\right|, t=nT. 
\label{eq11}
\end{equation}

This expression makes clear that a long $T_{2,eff}^\ast$ does not by itself imply improved dc sensitivity. To examine this distinction experimentally, we extract  $|\partial S / \partial \Delta |$ from the measured detuning-dependent signals and present the resulting response map in Fig. \ref{fig5}. The original measurements use a fixed projection phase, $\phi$. To remove the resulting oscillatory dependence, we reconstruct the phase-optimized response from the fitted fringe amplitude. For Ramsey, $|\partial S/ \partial \Delta|_{\mathrm{opt, R}} = 2\pi t A_{\mathrm{R}}(t)$, which scales as $te^{-t/T_2^*}$ and is maximal at $t = T_2^*$ for the single-exponential envelope used here. For WAHUHA, $|\partial S/ \partial \Delta|_{\mathrm{opt,W}} = 2\pi t A_{\mathrm{W}}(t, \Delta)|df_p/d\Delta|$,where $df_p/d\Delta$ is obtained from a local fit to the experimental Floquet spectral branch. While the raw response spectra in Fig. \ref{fig5} (a,b,c) exhibit oscillations from coherent beating among the hyperfine and the unoptimized readout phase, Fig. \ref{fig5} (d) shows a guide obtained by summing the maximum response of each manifold. This suppresses isotope-specific beating and provides a generalized benchmark across different nuclear spin environments.

\section{Conclusion and outlook}\label{sec4}

We have investigated the stroboscopic dynamics of a dense diamond NV ensemble under WAHUHA control using detuning-resolved spectroscopy and finite-pulse Floquet analysis. WAHUHA extends the observed stroboscopic decay time from $T_2^* \approx 0.9\ \mu$s to an effective inhomogeneous dephasing time of $T_{2,eff}^* \approx 31\ \mu$s. However, this thirtyfold increase produces little corresponding improvement in dc magnetic-field sensitivity. This result shows that, in periodically driven spin ensembles, a long-lived stroboscopic signal should not be automatically interpreted as metrologically useful coherence. 

The finite-pulse Floquet spectrum explains this discrepancy. At short interpulse delays, the one-cycle phase is approximately linear in detuning but contains finite-pulse-induced offsets and a small avoided-crossing-like gap. At longer delays, the accumulated phase approaches the $\pi$ phase boundary and folds back into the finite stroboscopic frequency window. The resulting phase wrapping and quasi-energy branch folding generate long-lived stroboscopic oscillations, including a $2T$ response, but simultaneously reduce the detuning-to-phase slope $d\Phi/d\Delta$. Thus, the long $T_{2,eff}^\ast$ cannot result in strong dc-field response.

A sensitivity improvement requires the increase in usable interrogation time to compensate for reduced dc transduction, finite-pulse overhead, and contrast loss. In the ideal equal-dwell limit, the $1/\sqrt{3}$ transduction factor implies that a threefold increase in the usable decay time is required only to reach the Ramsey sensitivity, and a larger increase is required for improvement. Possible extensions include unbalanced WAHUHA, which retains a larger dc response at the cost of incomplete interaction suppression, and a qutrit-based double-quantum extension. The latter uses full spin-1 control to access the $|+1\rangle \leftrightarrow |-1\rangle $ coherence, which accumulates dc magnetic phase at twice the single-quantum rate. Neither approach guarantees improved sensitivity, and more general Hamiltonian-engineering protocols may be required to suppress residual interactions while preserving dc transduction.

In summary, our results demonstrate that an extended fitted stroboscopic decay time under periodic driving does not necessarily yield a proportional improvement in dc magnetic-field sensitivity. This disconnect is not unique to WAHUHA or NV ensembles, but can arise more generally in periodically driven quantum sensors whenever the Floquet quasienergy structure produces a long-lived sampled response without preserving the corresponding parameter-to-signal transduction. The same consideration is relevant to other dc-sensitive dynamical-decoupling protocols \cite{kitamura2025} and to other driven spin or pseudospin platforms, including solid-state defect ensembles, trapped-ion systems, and cold-atom ensembles. Such protocols should therefore be benchmarked using the full measurable response-including contrast, decay envelope, readout noise, and Floquet transduction slope-rather than the fitted stroboscopic lifetime alone.








\section*{List of abbreviations}
NV, Nitrogen-vacancy; WAHUHA,  Waugh-Huber-Haeberle; NMR, Nuclear Magnetic Resonance; EPR, Electron Paramagnetic Resonance; CPMG, Carr-Purcell-Meiboom-Gill; AHT, Average Hamiltonian Theory; SBD, Spin-bath Driving; DEER,  Double Electron–Electron Resonance

\section*{Declarations}

\subsection*{Funding}
This work was supported by the Institute of Information \& Communications Technology Planning \& Evaluation (IITP) grant funded by the Korea government (MSIT) (No. RS-2025-02215576, RS-2025-25464252) and by the National Research Foundation (NRF) of Korea grant funded by the Korean government (MSIT) (No. RS-2024-00442710),  and Global-Learning \& Academic research institution for Master’s PhD students and Postdocs (G-LAMP) Program of the National Research Foundation of Korea (NRF) grant funded by the Ministry of Education (No.RS-2023-00285390) and Korea Research Institute of Standards and Science (Grant No. GP GP2024-0013).

\subsection*{Competing interests}
The authors declare that they have no competing interests

\subsection*{Availability of data and materials}
All data generated or analysed during this study are included in this published article.

\subsection*{Authors' contributions}
M.-C.N., U.K., S.-J.L., H.K., H.S., and S.O. wrote the main manuscript text and M.-C.N. prepared all figures. All authors reviewed and approved the final version of the manuscript.
\noindent






\begin{appendices}

\section{\label{sec:AppA}Average-Hamiltonian Theory and Finite-Pulse Floquet Model}

\begin{table}[ht]
	\caption{Principal symbols used in the Floquet analysis. }\label{tab1}
	\begin{tabular}{l|c|c}
		\toprule%
		\textbf{Symbols} & \textbf{Physical Description} & \textbf{Units}  \\ 
		\midrule
		$\tau$ & Free evolution interpulse delay between WAHUHA pulses & (ns) \\ \midrule
		$t_p$ & Duration of each WAHUHA $\pi/2$ pulse & (ns) \\  \midrule
		$T$ & Duration of one WAHUHA cycle, $T=6\tau+4t_p$ & (ns) \\  \midrule
		$T_2^\ast$ & Ramsey inhomogeneous dephasing time & ($\mu$s) \\  \midrule
		$T_{2,eff}^\ast$ & Phenomenological decay constant of the  & ($\mu$s) \\
		&stroboscopic WAHUHA envelope& \\ \midrule
		$\Delta$& Effective microwave detuning;& (MHz) \\
		&used a calibrated proxy for a dc-field shift & \\ \midrule
		$\Delta_{cross}$ & Detuning at which the longitudinal phase & (MHz) \\ 
		& contribution vanishes, $\Phi_0+2\pi \kappa \Delta T=0$ & \\ \midrule
		$\Delta_{\pi}$ & A Nyquist-boundary detuning  & (MHz) \\ \midrule
		$\Phi$&One-cycle Floquet eigenphase splitting& (rad) \\ \midrule
		$\Phi_{wr}$&Wrapped one-cycle Floquet phase& $(-\pi,\pi]$ (rad) \\ \midrule
		$\Phi_{c}$& One-cycle Floquet phase at branch cut& (rad) \\ \midrule
		$\Phi_{gap}$&Residual transverse Floquet components& (rad)\\ 
		&that sets the minimum eigenphase splitting& \\ \midrule
		$f_p$&Stroboscopic Fourier-peak frequency, $f_p=\Phi_{wr}/(2\pi T)$ & (MHz) \\ \midrule
		$S[n]$&Stroboscopic signal measured after $n$ WAHUHA cycles & (a.u.)\\ \midrule
		$\eta_B$&Dc magnetic-field sensitivity& (nT/$\sqrt{\text{Hz}}$)\\ 
		\botrule
	\end{tabular}
\end{table}

\subsection{\label{app:EffHamiltonian}Effective Hamiltonian of the dense NV ensemble}

The ground state of an NV center is a spin-1 system with basis \{$|+1\rangle,|0\rangle,|-1\rangle$\}. In the experiment, the zero-field splitting separates $|0\rangle$ from the $|\pm1\rangle$ manifold, while the axial bias field $B_z \approx$ 6 mT  separates the $|0\rangle\leftrightarrow |-1\rangle$ and $|0\rangle\leftrightarrow |+1\rangle$ transitions by approximately 330 MHz. This separation is much larger than the experimental Rabi frequency, $\Omega_R / 2 \pi \approx$ 25 MHz, and the relevant interaction scales. The $|+1\rangle$ state can be treated as an off-resonant spectator.

We define the projector onto the two-level manifold as $P=|0\rangle \langle0|+|-1\rangle \langle-1|$  and introduce effective spin-1/2 operators $s_\alpha=\sigma_\alpha/2$ in the ordered basis \{$|0\rangle,|-1\rangle$\}. To distinguish the two operator spaces, $\mathcal{S}_\alpha$ denotes a spin-1 operator, whereas $s_\alpha$ denotes its projected spin representation. Direct projection gives,
 
 \begin{equation}
 	P\mathcal{S}_x P=\sqrt{2}s_x, P\mathcal{S}_y P=\sqrt{2}s_y, P\mathcal{S}_z P=s_z-12/I.
 	\label{eqa1}
 \end{equation}
 
The WAHUHA pulses implement SU(2) rotations within the effective spin-1/2 manifold.
 
Starting from the secular NV-NV dipolar Hamiltonian,
 
\begin{equation}
 H_{dd}=\sum_{i<j}{D_{ij}\left( 2\mathcal{S}_{z,i}\mathcal{S}_{z,j}-\mathcal{S}_{x,i}\mathcal{S}_{x,j}-\mathcal{S}_{y,i}\mathcal{S}_{y,j}\right)}, 
 \label{eqa2}
\end{equation}
projection onto the selected manifold gives, 

\begin{equation}
	H_{NV-NV}^{eff}=2\sum_{i<j}{D_{ij}\left( s_{z,i}s_{z,j}-s_{x,i}s_{x,j}-s_{y,i}s_{y,j}\right)}.
	\label{eqa3}
\end{equation}

Equal-axis WAHUHA averaging then yields

\begin{equation}
\bar{H}_{NV-NV}^{0}=-\frac{2}{3}\sum_{i<j}{D_{ij}\mathbf{s}_i \cdot \mathbf{s}_j}.
	\label{eqa4}
\end{equation}

Thus, the projected interaction is not identically canceled. Instead, WAHUHA removes its anisotropic component while retaining an isotropic Heisenberg term. This residual term commutes with the collective effective spin and does not diphase a uniformly polarized state at zeroth order. The same projection reduces the longitudinal NV-P1 interaction to 
\begin{equation}
	H_{NV-P1}^{eff}=\sum_{i,j}{A_{i,k}S_{z,i}I_{z,k}}, 
	\label{eqa5}
\end{equation}
up to a P1-only frequency shift. This interaction is not fully canceled by WAHUHA and is further suppressed experimentally using spin-bath driving. All subsequent AHT and finite-pulse Floquet calculations are formulated within this effective spin-1/2 basis.

\subsection{\label{app:ValidityAHT}Range of validity of AHT and the finite-pulse Floquet description}
Average Hamiltonian theory and Floquet theory provide complementary descriptions of periodically driven dynamics. For the one-cycle propagator in Eq. (\ref{eq2}), $U_F (T)=\exp{(-iH_F T)}$,  the Floquet Hamiltonian can be expanded as
	
\begin{equation}
	H_F=\bar{H}^{(0)}+\bar{H}^{(1)}+\bar{H}^{(2)} +\cdots. 
	\label{eqa6}
\end{equation}	

The conventional ideal-pulse WAHUHA description corresponds to retaining only the zeroth-order term, $\bar{H}^{(0)}$. In this limit, $\bar{H}^{(0)}$ reduces to the dwell-time-weighted average of the Hamiltonians associated with the different WAHUHA intervals. 
The zeroth-order approximation is reliable when the evolution accumulated during each interpulse interval is small. A practical condition is

\begin{equation}
	\tau \ll W^{-1},J^{-1},
	\label{eqa7}
\end{equation}

where W and J are representative angular-frequency scales of the on-site disorder and spin-spin interaction, respectively \cite{Zhao2020}. As $\tau$ increases, higher-order terms in the Floquet Hamiltonian are no longer negligible, and the zeroth-order AHT approximation loses quantitative accuracy.
Finite pulse duration can also be incorporated within AHT as additional pulse-interval contributions \cite{Choi2020}. In the present work, however, we evaluate the one-cycle unitary generated by $H_{pulse}(t)$ directly as an ordered product of the free-evolution and finite-pulse propagators. This finite-pulse Floquet calculation does not rely on the short-$\tau$ approximation and is therefore applicable over the full range of interpulse delays considered here, within the effective two-level model.
At short interpulse delays, the Floquet eigenphase obtained from $U(T)$ is approximately linear in detuning and agrees with the leading-order AHT picture. At longer delays, the accumulated phase can cross the principal Floquet-zone boundary. Because the eigenphases of $U(T)$ are defined modulo-2$\pi$, the corresponding quasienergy branch is then folded back into the finite stroboscopic frequency window. This phase wrapping is distinct from the loss of accuracy of zeroth-order AHT. The former is a property of the exact Floquet eigenphases, whereas the latter results from neglecting higher-order and finite-pulse contributions.

\begin{equation}
\bar{H}^{(0)}=\frac{1}{T} \int_{0}^{T} \tilde{H}(t)dt,
\label{eqa8}
\end{equation}
and	
\begin{equation}
	\bar{H}^{(1)}=-\frac{i}{2T} \int_{0}^{T}dt_1\int_{0}^{t_1}dt_2\left[\tilde{H}(t_1),\tilde{H}(t_2)\right].
	\label{eqa9}
\end{equation}

The conventional ideal-pulse WAHUHA picture corresponds to retaining only the zeroth-order term, $\bar{H}^{(0)}$. For rapid, piecewise-constant control, this term reduces to the weighted average of the toggling-frame Hamiltonians. The leading-order approximation is reliable when the interpulse delay is short compared with the disrorder and interaction-induced evolution time scale, $\tau \ll 1/W,1/J$ \cite{Zhao2020}. As $\tau$ increases, the commutator contribution $\bar{H}^{(1)}T \sim O[(\Lambda T)^2]$ and higher-order terms are no longer negligible. Finite pulse widths can also be incorporated within AHT as additional pulse-interval contributions \cite{Choi2020}. In the ideal-pulse zeroth-order treatment of Sec. \ref{app:AHT}, these contributions are neglected, whereas they are retained directly in the ordered-product finite-pulse Floquet calculation of Sec. \ref{app:FloquetUnitary}

\subsection{\label{app:AHT}Zeroth-order WAHUHA average Hamiltonian}

The WAHUHA sequence consists of four $\pi/2$ pulses separated by fre-evolution intervals,
\begin{equation*}
\tau-X_{-\frac{\pi}{2}}-\tau-Y_\frac{\pi}{2}-2\tau-Y_{-\frac{\pi}{2}}-\tau-X_\frac{\pi}{2}-\tau. 
\end{equation*}

In the ideal-pulse limit, the sequence cyclically rotates the spin quantization axis through three orthogonal directions. In the toggling frame, the secular $S_z$ operator is mapped successively onto $S_x$, $S_y$, and $S_z$. Averaging over one WAHUHA cycle gives
\begin{equation}
{\bar{S}}_z^{(0)}=\frac{1}{3}(S_x+S_y+S_z). 
\label{eqa10}
\end{equation}

For the homonuclear dipolar interactions, the corresponding zeroth-order average Hamiltonian is 
\begin{equation}
{\bar{H}}_{NV-NV}^{(0)}=\frac{1}{3}\sum_{i<j}{D_{ij}\left[\left(3S_{xi}S_{xj}-\mathbf{S}_i{\cdot\mathbf{S}}_j\right)+\left(3S_{yi}S_{yj}-\mathbf{S}_i{\cdot\mathbf{S}}_j\right)+\left(3S_{zi}S_{zj}-\mathbf{S}_i{\cdot\mathbf{S}}_j\right)\right]}.
\label{eqa11}
\end{equation}
Using $S_{xi}S_{xj}+S_{yi}S_{yj}+S_{zi}S_{zj}=\mathbf{S}_i{\cdot\mathbf{S}}_j$, we obtain ${\bar{H}}_{NV-NV}^{(0)}=0$.

Thus, WAHUHA cancels the homonuclear dipolar interaction to leading order in the ideal-pulse limit. Importantly, the static detuning term is not eliminated. Instead,
\begin{equation}
\Delta S_z\rightarrow\frac{\Delta}{3}(S_x+S_y+S_z).
\label{eqa12}
\end{equation}

Therefore, unlike conventional echo-based dynamical decoupling, WAHUHA can suppress homonuclear dipolar interactions while retaining a finite dc-sensing channel. This is the reason that WAHUHA is appropriate for dc sensing, whereas conventional CPMG or XY-type sequences would refocus the phase from the target dc field.

\subsection{\label{app:NV-P1}Residual NV-P1 interaction and spin-bath driving}

Because the WAHUHA microwave pulses act only on the NV spins, P1 spins are not toggled by the sequence. The secular NV-P1 interaction therefore averages to 
\begin{equation}
{\bar{H}}_{NV-P1}^{(0)}=\frac{1}{3}\sum_{i,k}{A_{ik}\left(S_{x,i}+S_{y,i}+S_{z,i}\right)I_{z,k}}.
\label{eqa13}
\end{equation}

This term is reduced and tilted but not fully canceled. To further suppress this residual heteronuclear-like dephasing channel, an auxiliary spin-bath drive can be applied to the P1 spins. In the interaction picture of a bath drive with Rabi frequency $\Omega_{P1}$, the P1 operator evolves as
\begin{equation}
I_{z,k}(t)=I_{z,k}\cos{\left(\Omega_{P1}t\right)}-I_{y,k}\sin{\left(\Omega_{P1}t\right)}.
\label{eqa14}
\end{equation}

If  $\Omega_{P1}\gg A_{ik}$, the rapidly oscillating terms average to zero, giving
\begin{equation}
{\bar{H}}_{NV-P1,bath}^{(0)}\approx0.
\label{eqa15}
\end{equation}

This average-Hamiltonian picture provides useful intuition for why WAHUHA plus bath driving can extend the observed decay envelope. However, it does not determine the finite-pulse stroboscopic phase structure that controls the measured spectra and dc sensitivity.

\subsection{\label{app:FloquetUnitary} Exact finite-pulse Floquet unitary}

In the experiment, the WAHUHA cycle contains finite-width microwave pulses. The true cycle period is 
\begin{equation}
T=6\tau+4t_p,
\label{eqa16}
\end{equation}
where $t_p$ is the $\frac{\pi}{2}$-pulse duration. During each free-evolution or pulse segment, the Hamiltonian is piecewise constant. The unitary for a segment $k$ is 
\begin{equation}
U_k=\ e^{-iH_kt_k},
\label{eqa17}
\end{equation}
with $H_k=\frac{1}{2}\left[\Omega_{x,k}\sigma_x+\Omega_{y,k}\sigma_y+\left(\Delta+A_{zz}m_I\right)\sigma_z\right]. $

The one-cycle unitary is the ordered product
\begin{equation}
U(T,\Delta)=U_NU_{N-1}\cdots U_2U_1.
\label{eqa18}
\end{equation}

Equivalently, 
\begin{equation}
U\left(T,\Delta\right)=\mathcal{T} \exp{\left[-\int_{0}^{T}H\left(t,\Delta\right)\right]}=\exp\left[-iH_F(\Delta)T\right],
\label{eqa19}
\end{equation}
where $H_F$ is the Floquet Hamiltonian. For a two-level system,
\begin{equation}
H_F\left(\Delta\right)=\frac{1}{2}\mathbf{w}(\Delta)\cdot\boldsymbol{\sigma} .
\label{eqa20}
\end{equation}

The Floquet eigenphase splitting accumulated over one cycle is 
\begin{equation}
\Phi\left(\Delta,\tau\right)=|\mathbf{w}\left(\Delta,\tau\right)|T.
\label{eqa21}
\end{equation}

To extract $\Phi$ from the cycle unitary, we remote the global phase,
\begin{equation}
V=\ \frac{U(T,\Delta)}{\sqrt{\det{U(T,\Delta)}}},
\label{eqa22}
\end{equation}
 so that $V\in SU(2)$. Then 
\begin{equation}
V=\cos{\left(\frac{\Phi}{2}\right)I-i\sin{\left(\frac{\Phi}{2}\right)}\ }\hat{\mathbf{n}}\cdot\boldsymbol{\sigma},
\label{eqa23}
\end{equation}
 and the phase is obtained from 
\begin{equation}
\Phi\left(\Delta,\tau\right)=2\arccos \left \{\frac{\operatorname{Re} \left[ \operatorname{Tr}(V)\right]}{2}\right\}.
\label{eqa24}
\end{equation}

This expression is used throughout the paper to compute the finite-pulse Floquet spectrum.

\subsection{\label{app:Crossing}Phase wrapping and quasienergy folding}

The stroboscopic signal after $n$ WAHUHA cycles is modeled as 
\begin{equation}
S\left[n\right]=S_{off}+C\left(\tau\right)e^{-t/{T_{2,eff}^\ast}}\cos{\left[n\Phi_{wr}\left(\Delta,\tau\right)+\phi_r\right]},
\label{eqa25}
\end{equation}
and phase $\Phi_{wr}\in\left(-\pi,\pi\right]$ is the wrapped Floquet phase. The observed Fourier peak is
\begin{equation}
f_p(\Delta,\tau)=\frac{\Phi_{wr}(\Delta,\tau)}{2\pi T}.
\label{eqa26}
\end{equation}
Here, we assume error-free control. For a detailed analysis of control errors - including residual transverse Floquet components arising from finite pulse widths, pulse distortions, or off-axis control errors - see Appendix \ref{app:Gap}. Therefore, In the weak-phase accumulation regime, 
\begin{equation}
\Phi\left(\Delta,\tau\right)=\Phi_0(\tau)+2\pi\kappa(\tau)\Delta T.
\label{eqa27}
\end{equation}
The wrapped phase satisfies $\Phi_{wr}=\Phi-2\pi m,\ m\in\mathbb{Z}$. The zero-frequency crossings occur when $\Phi_{wr}=0$, or equivalently $\Phi=2\pi m$. Thus,

\begin{equation}
\Delta_{cross}^{\left(m\right)}\left(\tau\right)=\frac{2\pi m-\Phi_0}{2\pi \kappa T}=-\frac{\Phi_0}{2\pi\kappa T}+\frac{m}{\kappa T}.
\label{eqa28}
\end{equation}

The m=0 branch gives the primary crossing,
\begin{equation}
\Delta_{cross}^{\left(0\right)}\left(\tau\right)=-\frac{\Phi_0}{2\pi \kappa T}.
\label{eqa29}
\end{equation}
The additional crossings at larger $\tau$ arise from different wrapped branches of the same underlying Floquet dispersion, not from new physical resonances. Phase wrapping itself occurs when $\Phi=(2m+1)\pi$, so that $\Delta_{wr}^{\left(m\right)}\left(\tau\right)=\frac{(2m+1)\pi-\Phi_0}{2\pi \kappa T}$. At this boundary, the observed stroboscopic frequency reaches the Nyquist edge, $|f_p|=1/2T$.

\subsection{\label{app:Gap}Minimal finite-gap phenomenology}

The short-delay spectra shows both a finite detuning-axis offset and a small avoided-crossing-like gap. These features can be captured by a minimal finite-pulse Floquet model,
\begin{equation}
\Phi\left(\Delta,\tau\right)=\sqrt{(\Phi_0+2\pi\kappa\Delta T)^2+\Phi_{gap}^2}.
\label{eqa30}
\end{equation}
Here, $\Phi_0$ is a longitudinal cycle-phase intercept, while $\Phi_{gap}$ parameterizes a residual transverse Floquet component arising from finite pulse width, pulse distortion, or off-axis control errors. In this model, $\Phi_0$ shifts the zero-frequency crossing, while $\Phi_{gap}$ prevents the eigenphase splitting from vanishing at the crossing. The center of the avoided crossing is determined by $\Phi_0+2\pi\kappa\Delta T=0$, which gives $\Delta_{cross}\left(\tau\right)=-\frac{\Phi_0}{2\pi \kappa T}$. Thus the detuning-axis offset decreases approximately as $1/T$, consistent with the experimentally observed return of the crossing toward zero detuning as $\tau$ increases.

\subsection{\label{app:Nyquist}Quadratic Floquet dispersion near the Nyquist boundary }
Eq. (\ref{eq6}) describes the weak-phase-accumulation regime in which the Floquet phase is approximately linear in detuning. To account for the curvature observed in the long-delay spectrum, we expand the Floquet branch near a Nyquist-boundary detuning $\Delta_{\pi}$, defined by
\begin{equation}
	\Phi\left(\Delta_{\pi},\tau\right)=P_m, P_m = (2m+1)\pi.
	\label{eqa31}
\end{equation}

Because the principal wrapped phase is discontinuous at the branch cut, we denote the corresponding phase by $\Phi_c$. Defining $\delta \Delta = \Delta-\Delta_{\pi}$, the local dispersion is

\begin{equation}
	\Phi_c(\Delta, \tau) \simeq P_m + 2\pi T \left[ a_1 \delta\Delta + \frac{a_2}{2} \delta\Delta^2 \right].
	\label{eqa32}
\end{equation}

Equivalently, the continuously tracked spectral branch is

\begin{equation}
	\Phi_c(\Delta, \tau) \simeq \frac{s}{2T}  a_1 \delta\Delta + \frac{a_2}{2} \delta\Delta^2, s=\pm{1},
	\label{eqa33}
\end{equation}

where $ a_1 = \left.\frac{\partial f_p}{\partial \Delta}\right|_{\Delta_\pi}, \quad a_2 = \left.\frac{\partial^2 f_p}{\partial \Delta^2}\right|_{\Delta_\pi}$.

It follows that 
\begin{equation}
\frac{1}{2\pi T} \frac{\partial \Phi_c}{\partial \Delta} = \frac{\partial f_p}{\partial \Delta} \simeq a_1 + a_2(\Delta - \Delta_\pi).
	\label{eqa34}
\end{equation}

Because $\Phi_c$ and $\Phi_{wr}$ differ only by an integer multiple of 2$\pi$, they produce the same stroboscopic signal. Using Eq. (\ref{eq4}), the maximum magnetic-field response at the optimal readout quadrature is

\begin{equation}
	\left| \frac{\partial S[n]}{\partial B} \right|_{\text{max}} = 2\pi nT \, C(\tau) e^{-nT / T_{2,\text{eff}}^*} \left| a_1 + a_2(\Delta - \Delta_\pi) \right| \left| \frac{\partial \Delta}{\partial B} \right|.
	\label{eqa35}
\end{equation}
The corresponding magnetic sensitivity is
 
 \begin{equation}
 	\eta_B^{\text{opt}} = \frac{\sigma_S \sqrt{t_m}}{2\pi nT \, C(\tau) e^{-nT/T_{2,\text{eff}}^*} \left| a_1 + a_2(\Delta - \Delta_\pi) \right| \left| \partial\Delta / \partial B \right|}.
 	\label{eqa36}
 \end{equation}
 
Above equations directly relate the local Floquet spectral slope to the dc magnetic-field sensitivity. In the short-delay linear regime, $a_1 \simeq \kappa(\tau)$ and $a_2 \simeq 0$. A fit to the data at $\tau$ = 12 ns gives $a_1=0.542(7)$ and $a_2=0$ (Fig. \ref{fig3} (b)). In the long-delay regime, the curved finite-pulse Floquet branch can exhibit a reduced but finite $a_1$, offsetting the sensitivity gain expected from an extended $T_{2,eff}^\ast$. The special case $a_1=0$ corresponds to a tangent turning point and hence a first-order magnetic blind spot, rather than enhanced sensitivity. The coefficients $a_1$ and $a_2$ are extracted directly from the simulated Floquet branch. At $\tau$ = 220 ns, fitting the data yields  $a_1=-0.73(2))$ and $a_2=0.366(9)$ (Fig. \ref{fig3} (d)).

Modulo-2$\pi$ wrapping does not alter the local phase derivative by itself. Any reduction in dc transduction must arise from the local dispersion of the finite-pulse Floquet branch in the folded regime and should be verified from the simulated branch.

\section{Experimental Details}
\subsection{\label{app:Setup}Wide-field NV ensemble microscope}
\begin{figure*}
\centering
\includegraphics[width=0.8\textwidth]{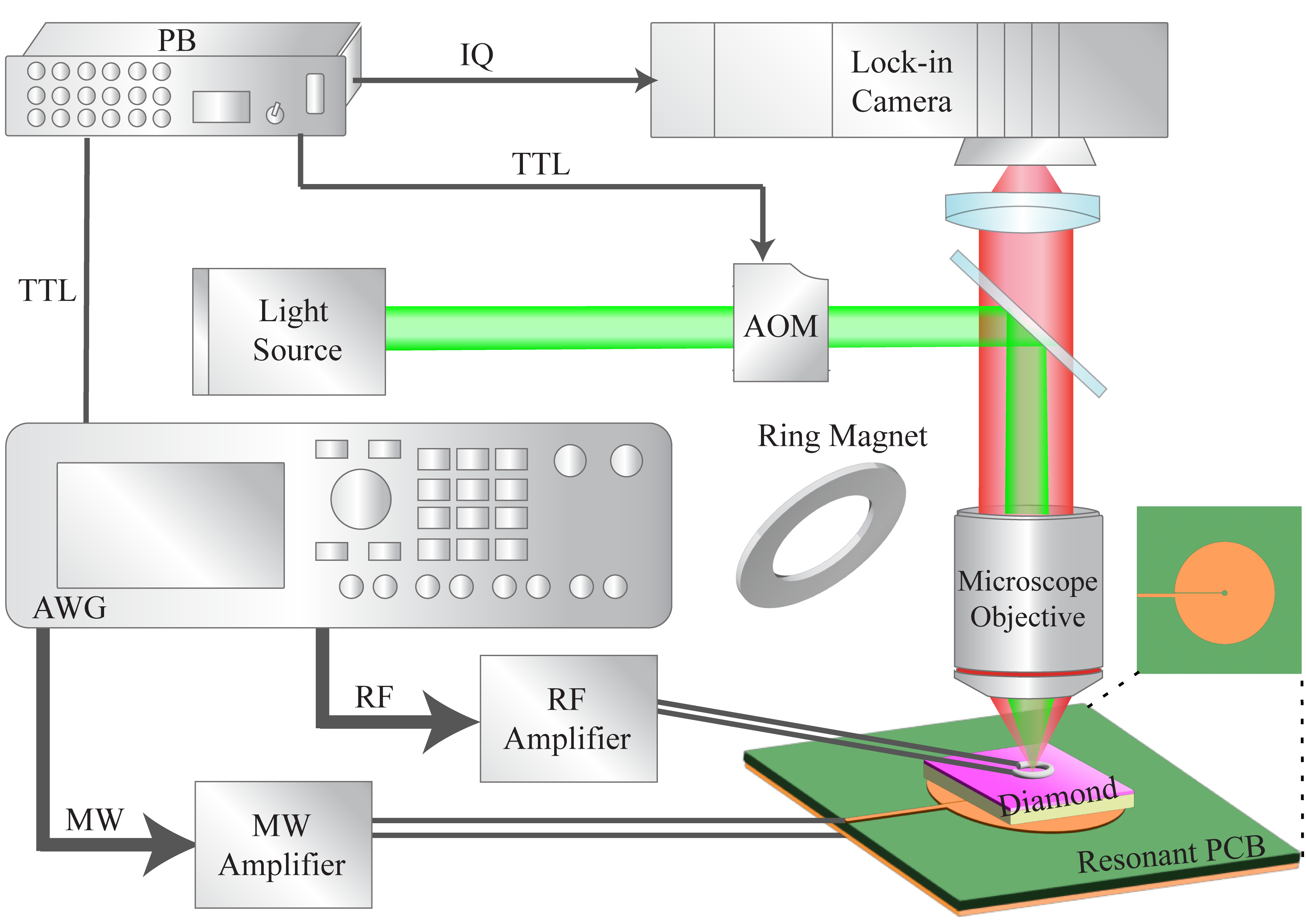}
\caption{\label{fig6}Schematic diagram of the experimental setup for wide-field imaging and coherent spin control of NV centers. A multi-channel PB acts as the master clock, dispatching TTL trigger signals to synchronize the AWG and the AOM, while supplying a reference signal to the lock-in camera. For optical spin initialization and readout, a continuous green light source is pulsed by the AOM. Spin-dependent photoluminescence from the NV centers is collected by the same objective and imaged onto the lock-in camera. Coherent spin manipulation is driven by MW pulses synthesized by the AWG, amplified, and delivered to the sample via a custom resonant PCB. The inset details the planar ring antenna geometry of the PCB used to generate a MW field. A static magnetic bias field is applied using a permanent ring magnet.}
\end{figure*}
The experiment was performed using a wide-field microscope designed for ensemble NV measurements. A green laser was used for optical spin initialization and readout. The laser was pulsed by an acousto-optic modulator (AOM G\&H, 3250-220), and spin-dependent photoluminescence from the NV ensemble was collected by the same objective (Nikon CFI Plan Apo Lambda D 20X) and imaged onto a lock-in camera (Heliotis, Helicam C3 ). A multi-channel pulse generator (Spincore, PBESR-PRO-500) provided the timing signal and synchronized the acousto-optic modulator, arbitrary waveform generator, microwave pulses, and camera acquisition. The current setup description lists an AOM, lock-in camera, pulse generator, AWG, microwave amplifier, resonant PCB, and ring magnet in the wide-field apparatus. 
Microwave control waveforms were synthesized by an arbitrary waveform generator and amplified before delivery to the diamond sample through a custom resonant printed circuit board. The microwave waveform defined the initial and final $\pi/2$ pulses, the WAHUHA pulse train, and the detuning used in the stroboscopic spectroscopy measurements. The WAHUHA pulse phases -X, Y, -Y, X and the interpulse delays were programmed directly in the waveform generator.
A static magnetic bias field was applied using a permanent ring magnet. The field was aligned close to the selected NV crystallographic orientation in order to isolate one $m_s = 0 \leftrightarrow m_s=-1$ transition for coherent control. In the Double Electron–Electron Resonance (DEER) measurements described below, the field magnitude was approximately 6 mT.

\begin{figure*}
	\centering
	\includegraphics[width=0.9\textwidth]{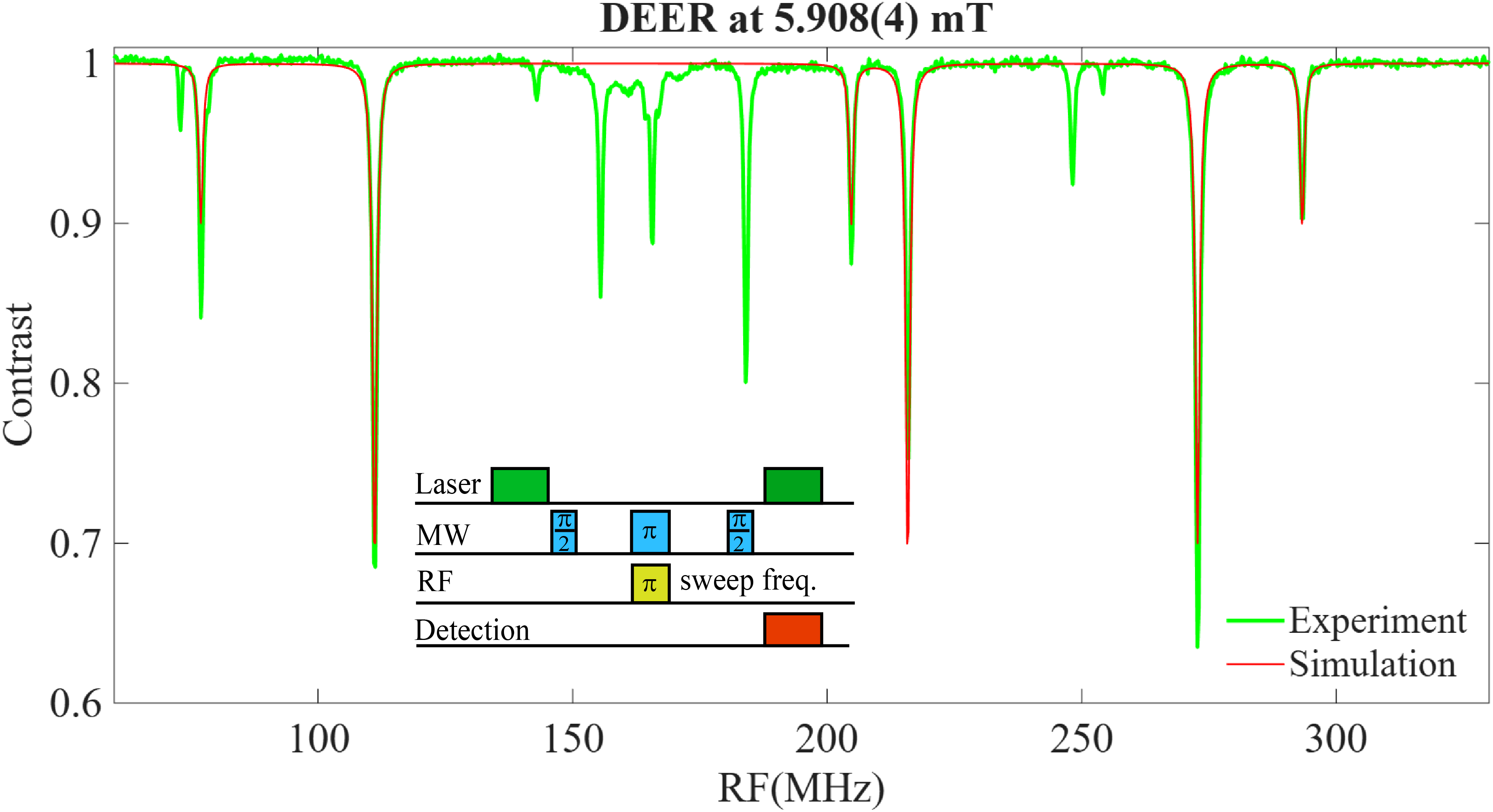}
	\caption{\label{fig7}DEER spectroscopy of the P1 spin bath. The figure shows the DEER spectrum at $B_0$ = 5.908(4) mT, with the magnetic field aligned along the [111] NV axis. The NV ensemble is detected using a microwave spin-echo sequence, while an RF $\pi$ pulse (150ns) is applied to the P1 bath during the echo free-evolution interval. The Rf frequency is swept from 60 MHz to 330 MHz in 0.25 MHz increments. Resonant RF excitation of the P1 bath produces dips in the NV echo contrast. The experimental spectrum is compared with a simulation to identify the P1 transitions used for spin-bath driving. }
\end{figure*}
\subsection{\label{app:Sample}Diamond sample}
The diamond sample, grown by ElementSix, contained a high-density near surface NV ensemble in a 4 µm-thick active layer. The initial substitutional nitrogen concentration was 12 ppm, and the resulting NV concentration was 3.6 ppm. The high NV density places the system in a regime where homonuclear NV-NV dipolar interactions and heteronuclear-like coupling to P1 centers strongly affect the dephasing dynamics. The diamond was isotopically purified, with a \textsuperscript{13}C concentration of 0.005\%, reducing the contribution from the spin-1/2 nuclear bath. 
Because the dominant remaining bath consists of electron spins, P1, an auxiliary spin-bath driving was used in selected measurements to suppress slow magnetic noise from the P1 bath. The DEER and Ramsey measurements described below characterize this bath and its effect on the NV ensemble $T_2^*$.

\subsection{\label{app:DEER}DEER spectroscopy of the P1 spin bath}

\begin{figure*}
	\centering
	\includegraphics[width=0.85\textwidth]{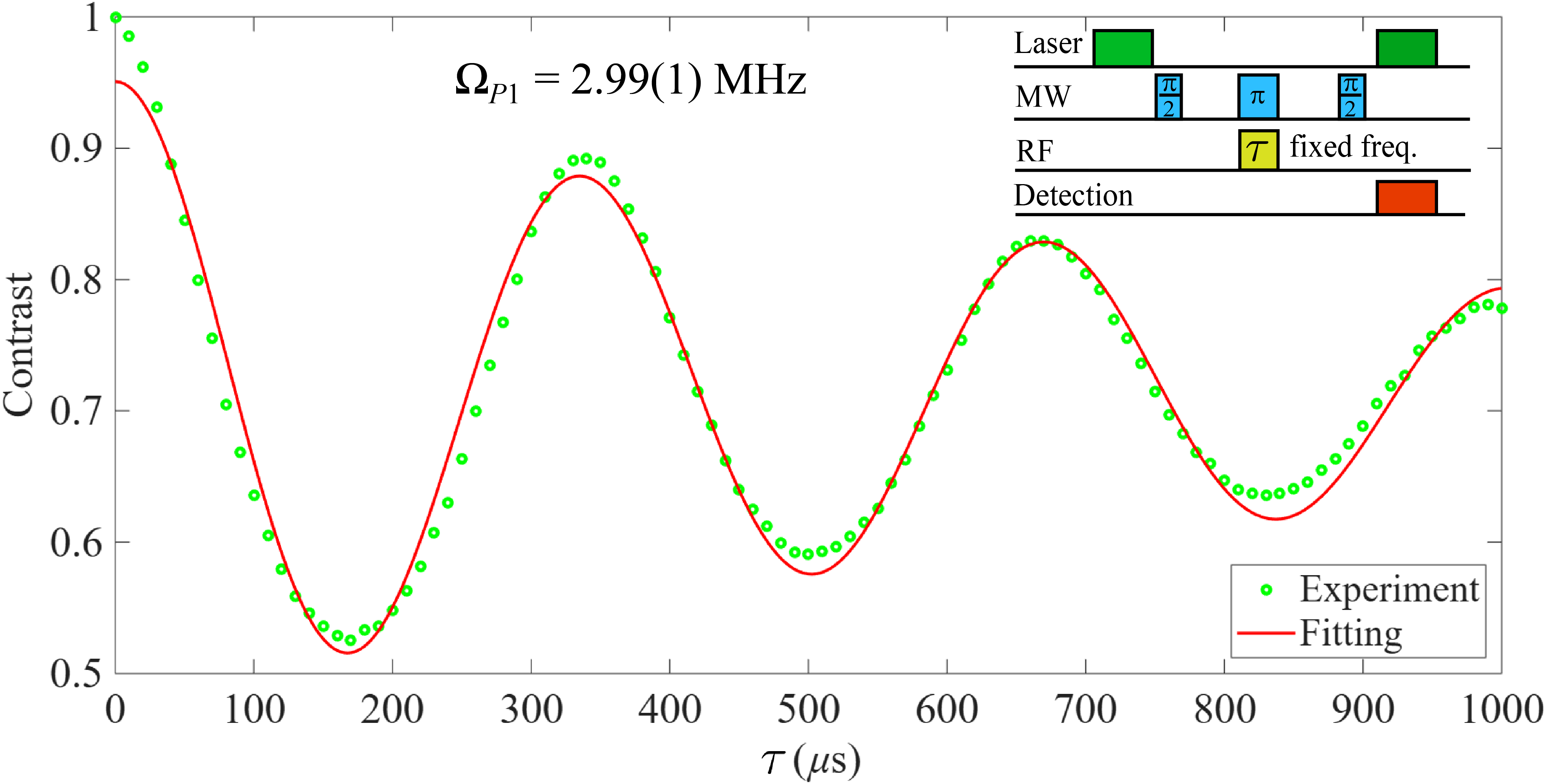}
	\caption{\label{fig8}P1 Rabi calibration for spin-bath driving. Rabi oscillations of the P1 spin bath are measured at the resonances identified from the DEER spectroscopy. The RF pulse duration is swept from 0 ns to 1000 ns in 10 ns increments. The solid curve represents a damped sinusoidal fit, yielding a P1 Rabi frequency of $\Omega_{P1}/2 \pi=2.99(1)$ MHz. This calibrated value is used to set the spin-bath driving strength in the Ramsey and WAHUHA measurements.}
\end{figure*}

To identify the P1 transition frequencies used for spin-bath driving, we performed DEER spectroscopy \cite{Knowles2014,Cooper2020,Rosenfeld2018} at $\mathbf{B}_0$ = 5.908(4) mT. The NV ensemble was used as a detector of the surrounding P1 spin-bath. The NV spins were manipulated using a microwave spin-echo sequence, ($\pi/2$ - $\pi$ - $\pi/2$), while an RF $\pi$ pulse with swept frequency was applied to the P1 bath during the free-evolution interval. When the RF frequency matched a P1 transition, the P1 spin state was inverted during the NV echo sequence. This sudden bath-spin flip changed the local magnetic field experienced by the NV ensemble and reduced the refocusing efficiency of the spin echo, producing a dip in the measured NV contrast. 
The resulting DEER spectrum showed multiple resonance dips as a function of RF frequency. These dips correspond to the hyperfine-split and orientation-dependent transition spectrum of P1. The measured spectrum was compared with a theoretical simulation, and the agreement allowed us to identify the P1 transition frequencies used for subsequent spin-bath driving.

\subsection{\label{app:SBD}P1 Rabi characterization and spin-bath driving}

After identifying the P1 resonance frequencies by DEER spectroscopy, we characterized the Rabi frequency of the applied spin-bath drive. The RF drive was tuned to a selected P1 transition, and the drive duration was varied while monitoring the NV response. The resulting oscillation was used to calibrate the bath-drive Rabi frequency $\mathrm{\Omega}_{P1}$.
In the experiments where spin-bath driving was applied, a continuous or quasi-continuous RF field was applied to the P1 transition during the NV evolution. The purpose of this drive was not to generate a coherent NV signal directly, but to rapidly modulate the P1 bath operators so that the residual NV-P1 coupling was averaged out. In the interaction picture of the bath drive, $I_z(t)=I_z\cos{\left(\Omega_{P1}t\right)}-I_y\sin{\left(\Omega_{P1}t\right)}$, so that the secular NV-P1 interaction averages to zero when $\Omega_{P1}$ is large compared with the NV-P1 coupling strength. This calibration therefore provides an experimental basis for using spin-bath driving to suppress the heteronuclear-like dephasing channel.

The uncertainties of the fit-derived parameters are reported as 95\% confidence intervals. Because the DEER and P1 Rabi measurements are calibration scans rather than noise measurements, photon shot noise and microwave phase noise are not independently resolved. Photon shot noise contributes to optical-signal fluctuations, whereas microwave phase may contribute to reduced contrast, spectral broadening, and systematic pulse-phase errors.

\subsection{\label{app:SBDDephasing}Ramsey \texorpdfstring{$T_{2}^{*}$}{T2*} with and without spin-bath driving}
\begin{figure*}
\centering
\includegraphics[width=0.85\textwidth]{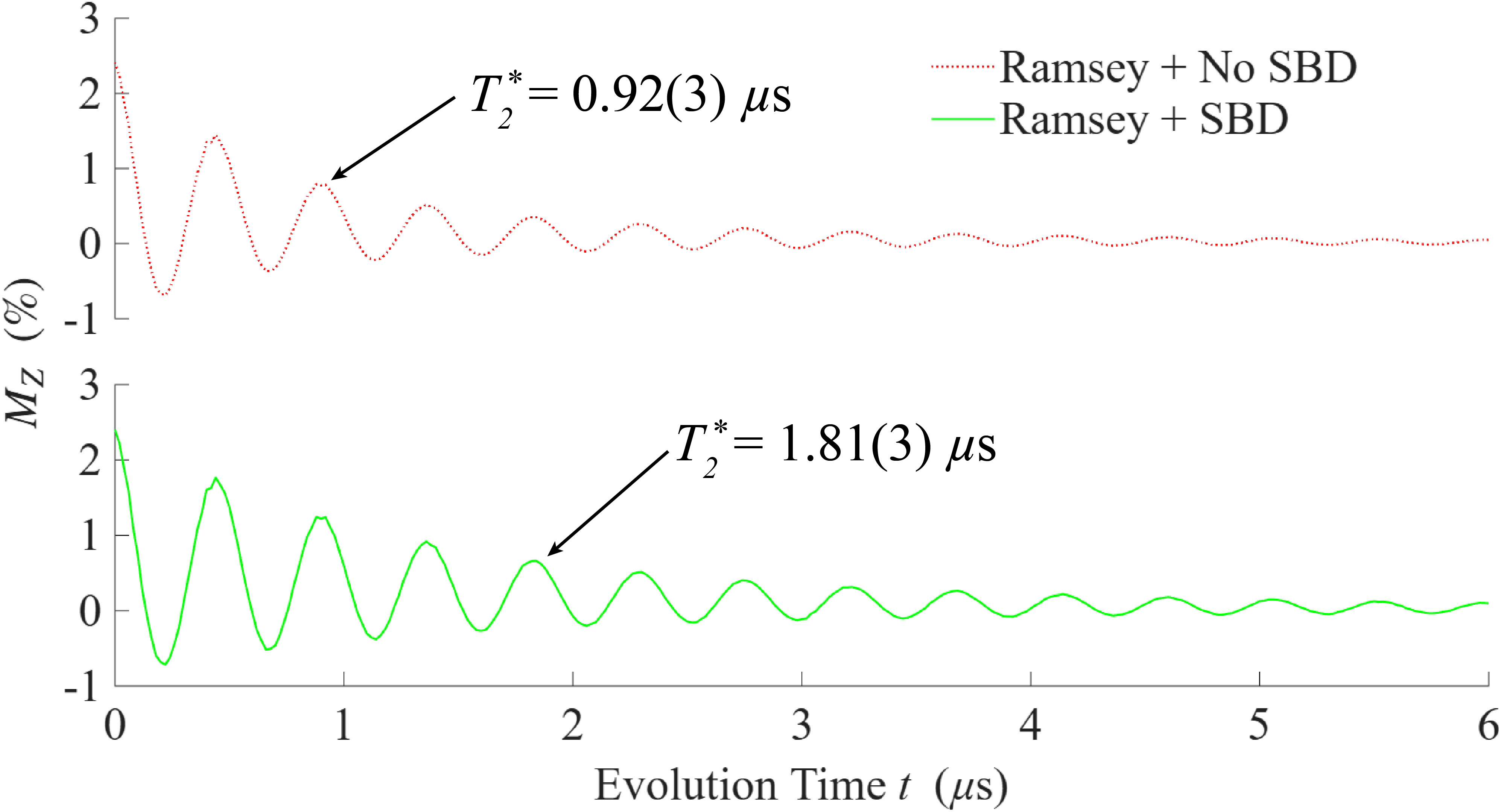}
\caption{\label{fig9}Ramsey inhomogeneous dephasing with and without spin-bath driving. Ramsey free-induction decay of the NV ensemble is measured with the P1 spin-bath drive off and on. The solid curves are fit to Eqs. (\ref{eqa37}) and (\ref{eqa38}), yielding $T_{2,off}^\ast=0.92(4) \mu$s and $T_{2,on}^\ast =1.81(3) \mu$s, respectively.}
\end{figure*}
To quantify the contribution of the P1 spin bath to the inhomogeneous dephasing of the NV ensemble, we measured Ramsey free-induction decay with and without spin-bath driving. The longitudinal magnetization, $M_z(t)$, without spin-bath driving was fit to 
\begin{equation}
M_z\left(t\right)=M_0+A\exp{\left(-\frac{t}{T_{2,off}^*}\right)}\cos(2\pi ft+\phi),
\label{eqa37}
\end{equation}
where $T_{2,off}^\ast$ is the inhomogeneous dephasing time in the absence of the bath drive. With spin-bath driving applied continuously to the selected P1 transition, the Ramsey signal was fit to the same functional form,
\begin{equation}
M_z\left(t\right)=M_0+A\exp{\left(-\frac{t}{T_{2,on}^*}\right)}\cos(2\pi ft+\phi),
\label{eqa38}
\end{equation}
where $T_{2,on}^\ast$ is the inhomogeneous dephasing time with P1 bath driving.
With spin-bath driving, the Ramsey inhomogeneous dephasing time increased from $T_{2,off}^\ast\approx 0.92\ \mu$s to $T_{2,on}^\ast\approx 1.8\ \mu$s, confirming that the P1 bath contributes significantly to the static dephasing environment.

\subsection{\label{app:SBD_WAHUHA}WAHUHA and spin-bath driving at short interpulse delay}	

\begin{figure*}
		\centering
		\includegraphics[width=0.75\textwidth]{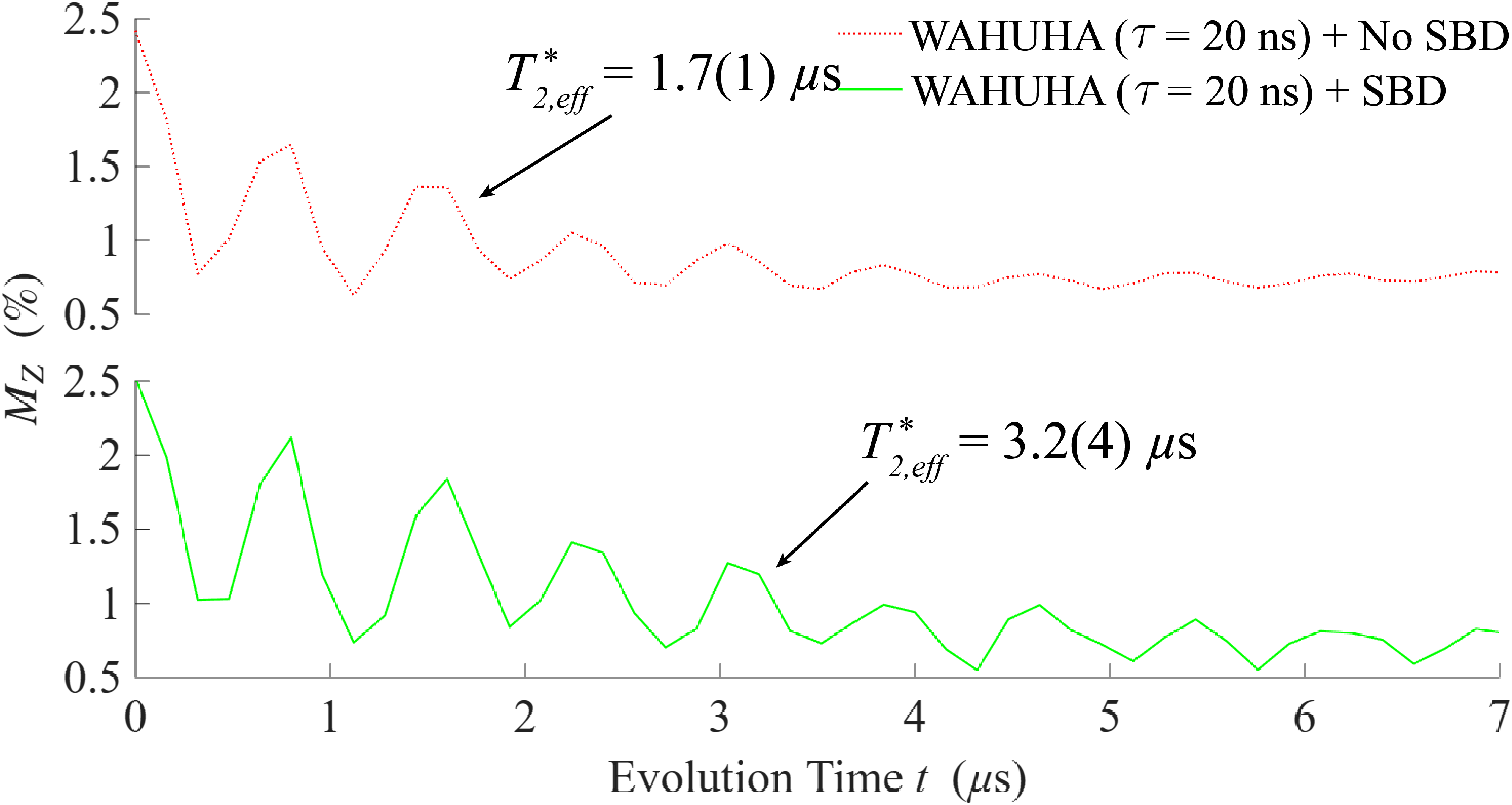}
		\caption{\label{fig10}Short-delay WAHUHA dynamics with and without spin-bath driving. Stroboscopic magnetization measured at $\tau$ = 20 ns without spin-bath driving (red) and with spin-bath driving(green). The fitted decay constants are $T_{2,eff}^{*}$ = 1.7(1) $\mu$s and $T_{2,eff}^{*}$ = 3.2(4) $\mu$s, respectively. The comparison indicates that residual coupling to the P1 bath substantially limits the short-delay WAHUHA envelope.}
	\end{figure*}

To determine whether the short-delay WAHUHA decay is limited by the residual P1 bath, we compare the stroboscopic dynamics at $\tau$ = 20 ns with and without spin-bath driving (Fig. \ref{fig10}). Without spin-bath driving, the fitted decay constant is $T_{2,eff}^{*}$ = 1.7(1) $\mu$s. When spin-bath driving is applied under identical WAHUHA conditions, the fitted value increases to $T_{2,eff}^{*}$ = 3.2(4) $\mu$s. This comparison shows that residual NV-P1 coupling is an important limitation of the short-delay WAHUHA response. WAHUHA and spin-bath driving suppress different interaction channels and provide complementary improvements. The combined enhancement is reported here as an empirical observation.

\subsection{\label{app:inter,ediate delay}Detuning dependence of the stroboscopic decay at an intermediate interpulse delay}
\begin{figure*}
	\centering
	\includegraphics[width=0.95\textwidth]{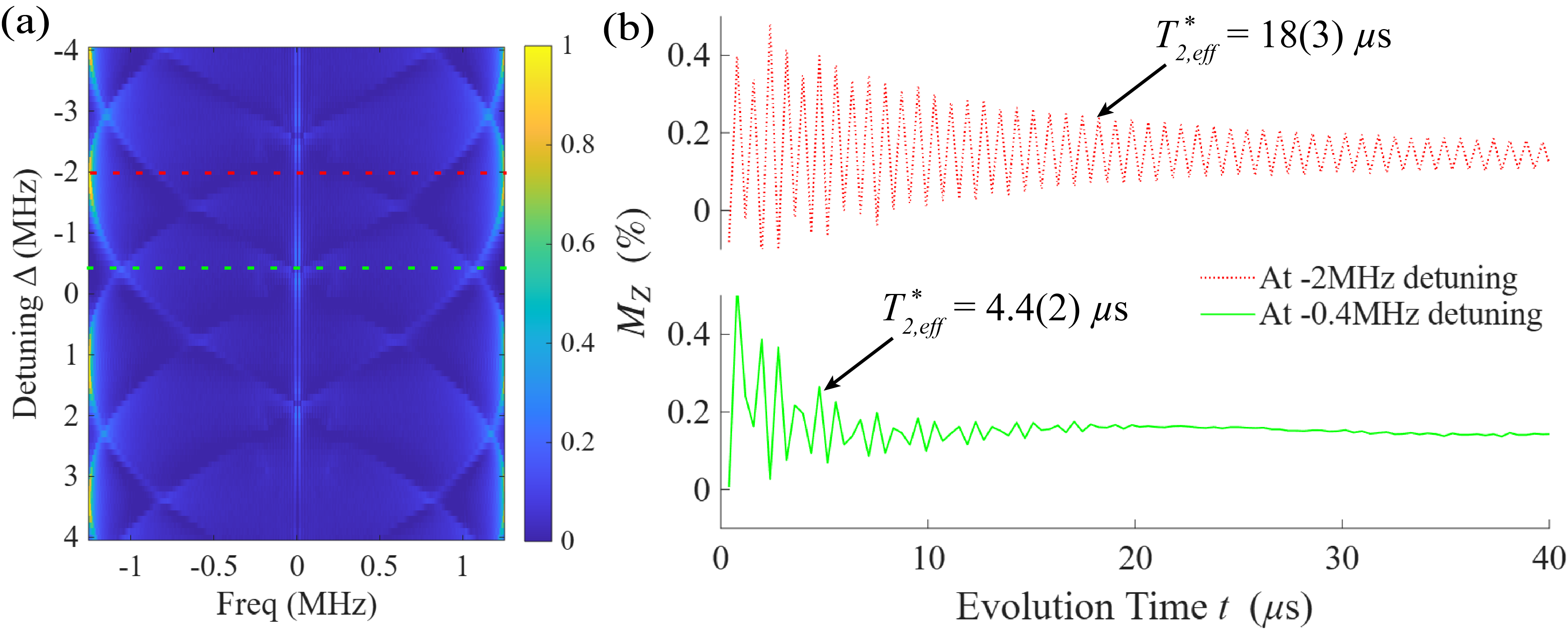}
	\caption{\label{fig11}Detuning dependence of the stroboscopic WAHUHA decay at $\tau$ = 60 ns. (a)Detuning-resolved stroboscopic spectrum. The red and green dashed lines indicate $\Delta$ = -2.0 MHz and -0.4 MHz, respectively. (b) Time-domain magnetization measured at the two selected detunings. The fitted stroboscopic decay constants are $T_{2,eff}^{*}$  = 18(3) $\mu$s at $\Delta$ = -2.0 MHz and $T_{2,eff}^{*}$  = 4.4(2) $\mu$s at $\Delta$ = -0.4 MHz. The comparison demonstrates that the fitted WAHUHA envelope lifetime depends strongly on the Floquet operating point even at a fixed interpulse delay.}
\end{figure*}

To provide an additional investigation at an intermediate interpulse delay, we measure the stroboscopic decay dynamics at $\tau$ = 60 ns for two detunings selected from the spectrum in Fig. \ref{fig11}(a). At $\Delta$ = -2.0 MHz, near a folded Floquet-branch region, the fit gives $T_{2,eff}^{*}$ = 18(3) $\mu$s.  At $\Delta$ = -0.4 MHz, within an approximately linear branch region, the fit gives $T_{2,eff}^{*}$ = 4.4(2) $\mu$s. Because the interpulse delay, pulse duration, pulse phases, and microwave amplitude are unchanged, the large difference between the two fitted decay constant demonstrates that $T_{2,eff}^{*}$ depends strongly on the selected Floquet branch. The effective decay time should therefore be regarded as a function of both interpulse delay and detuning, $T_{2,eff}^{*}(\tau,\Delta)$, rather than as a monotonic coherence time determined by $\tau$ alone. 

\begin{table}[ht]
	\caption{Summary of Experimental Parameters}\label{tab2}
	\begin{tabular}{l|c|c}
		\toprule%
		\textbf{Category} & \textbf{Parameter} & \textbf{Value}  \\ 
		\midrule
	 	& $[N_s^0]_{ini}$ & 12 ppm \\ \cmidrule{2-3}
	 	Sample & $[\text{NV}] = [\text{NV}^-] + [\text{NV}^0]$ & 3.6 ppm \\  \cmidrule{2-3}
	 	& $[{}^{13}\text{C}]$ & 0.005\% \\  \midrule
	\multirow{2}{*}{Static bias field} 
	 & $B_0$ & 5.908(4) mT \\  \cmidrule{2-3}
	& Orientation & Aligned along [111] NV axis \\ \midrule
	
	& Center frequency & 2.70216 GHz \\ \cmidrule{2-3}
	MW drive & $t_{\pi/2}$ & 10 ns \\  \cmidrule{2-3}
	& $\tau$ & Swept from 4 ns to 220 ns \\ \midrule
	
	\multirow{2}{*}{RF drive} 
	& RF Frequency Range & Swept 60 MHz to 330 MHz  (DEER) \\ \cmidrule{2-3}
	& P1 Drive Frequency & 77, 111, 205, 216, 273, 293.5  (MHz)\\
	\botrule
	\end{tabular}
\end{table}



\end{appendices}


\bibliography{references}

\end{document}